\documentclass[aps,prd,reprint,nofootinbib]{revtex4-2}

\usepackage{amsmath,amssymb}
\usepackage{graphicx}
\usepackage{booktabs}
\usepackage{hyperref}
\usepackage{xcolor}%

\begin{document}

\title{Astrophysical Constraints on Hadron--Quark Crossover in Hybrid Neutron Stars}

\author{G\'abor L\'aszl\'o Kasza}
\email{kasza.gabor@wigner.hun-ren.hu}
\affiliation{Theory Department, HUN-REN Wigner Research Centre for Physics, P.O. Box 49, Budapest H-1525, Hungary}
\affiliation{Institute of Technology, Hungarian University of Agriculture and Life Sciences (MATE), Károly Róbert Campus, Gyöngyös H-3200, Hungary}

\author{Gy\"orgy Wolf}
\email{wolf.gyorgy@wigner.hun-ren.hu}
\affiliation{Theory Department, HUN-REN Wigner Research Centre for Physics, P.O. Box 49, Budapest H-1525, Hungary}

\begin{abstract}
We investigate astrophysical constraints on hybrid neutron-star equations of state constructed by combining an extended linear sigma model description of quark matter
with several representative hadronic equations of state through a smooth hadron--quark crossover. Four different hadronic models are examined in order to explore the model
dependence of the resulting hybrid equations of state. After imposing stability, causality, perturbative QCD matching, tidal deformability, and maximum-mass constraints, only the eLSM+GMSR(BSK16) construction remains fully consistent with current astrophysical observations. The surviving parameterizations are further constrained using Bayesian inference based on NICER mass--radius measurements and the GW170817 gravitational-wave event. The posterior distributions favor a broad hadron--quark crossover occurring at relatively high baryon densities, while the formation of extended pure quark cores is strongly disfavored. Instead, the current observations favor neutron stars containing an extended mixed hadron--quark transition region. We also investigate the implications of the inferred maximum-mass distribution for the secondary object of GW190814 and find that, within the present model and adopted observational constraints, its interpretation as a neutron star is unlikely.
\end{abstract}

\maketitle

\section{Introduction}

Understanding the properties of strongly interacting matter under extreme conditions is one of the primary goals of modern nuclear physics. The phase structure of Quantum Chromodynamics (QCD) has been investigated extensively over the past decades using both theoretical and experimental approaches. At vanishing or small baryon chemical potential, lattice QCD calculations together with ultra-relativistic heavy-ion collisions have established many properties of hot strongly interacting matter, including the crossover transition to the quark--gluon plasma~\cite{Grelli:2017pkp,Sakaguchi:2019azf,Aparin:2023fml,Aoki:2006we,Borsanyi:2010cj,Borsanyi:2013bia}. At asymptotically large baryon densities ($\rho \sim 40\,\rho_0$, where $\rho_0$ is the nuclear saturation density), perturbative QCD (pQCD) provides reliable predictions for the equation of state (EoS)~\cite{Gorda:2022jvk}. Between these two regimes, however, the QCD phase diagram remains largely unexplored due to the sign problem of lattice QCD~\cite{Stephanov:2004wx,deForcrand:2009zkb,Nagata:2021ugx} and the limited experimental access to cold, dense matter.

Neutron stars (NSs) constitute the only known laboratories where such extreme conditions naturally occur. Central densities may exceed several times the nuclear saturation density while the temperature remains negligible compared to the relevant QCD energy scales. Consequently, observations of compact stars provide unique constraints on the cold EoS of strongly interacting matter. During the last decade, the rapidly increasing precision of NS observations has transformed this field into one of the most important probes of dense QCD.

The appearance of hyperons in dense NS matter gives rise to the so-called hyperon puzzle~\cite{Burgio:2021vgk,Logoteta:2021iuy}. Hyperons are expected to appear in dense matter once the chemical potential becomes sufficiently high. Their appearance generally softens the hadronic EoS and leading to lower maximum masses than those inferred from observations of massive pulsars~\cite{Logoteta:2021iuy,Vidana:2021ppe}. While many hyperonic EoSs therefore struggle to support NSs with masses around or above $2\,M_\odot$, this tension is strongly model dependent. Several relativistic approaches have demonstrated that sufficiently repulsive hyperonic interactions, particularly in the vector-meson sector, can support two-solar-mass NSs containing hyperons~\cite{Bednarek:2011gd,Weissenborn:2011kb,Weissenborn:2011ut,Lopes:2013cpa}. More recent Bayesian analyses incorporating astrophysical and hypernuclear constraints have likewise found hyperonic EoSs compatible with massive NSs and current multimessenger observations~\cite{Sun:2022yor,Malik:2022jqc}.
Beyond purely hadronic descriptions, the possible appearance of deconfined quark matter provides another scenario for the high-density interior of NSs. In particular, a smooth hadron--quark crossover can stiffen the high-density EoS without introducing a strong first-order phase transition~\cite{Masuda:2015kha, Baym:2017whm}.

Recent astrophysical observations provide increasingly stringent constraints on the EoS. Precision mass--radius measurements performed by the NICER mission have significantly reduced the allowed region of the mass--radius plane~\cite{Vinciguerra:2023qxq,Salmi:2024aum,Mauviard:2025dmd,Choudhury:2024xbk,Salmi:2024bss}, while the binary NS merger GW170817 established valuable constraints on the tidal deformability of NSs~\cite{LIGOScientific:2017vwq,LIGOScientific:2018hze}. At the same time, observations of massive pulsars through Shapiro-delay measurements indicate that the maximum mass of stable NSs must reach, or even exceed, approximately $2\,M_\odot$~\cite{Demorest:2010bx,Antoniadis:2013pzd,NANOGrav:2019jur}, with recent population analyses suggesting a $68\%$ credible lower limit of $M_{\rm TOV}>2.27\,M_\odot$~\cite{Romani:2022jhd,Romani:2025ytn}. Complementary multimessenger analyses based on EoS reconstruction have inferred $M_{\rm TOV}=2.25^{+0.08}_{-0.07}\,M_\odot$ by combining pulsar mass measurements, GW170817, NICER observations, and theoretical constraints from $\chi$EFT and pQCD~\cite{Fan:2023spm}.

In addition to these astrophysical constraints, the EoS must also be consistent with perturbative QCD calculations at asymptotically high baryon densities. Although the densities at which pQCD becomes quantitatively reliable are far beyond those realized in NSs, the EoS must smoothly approach this regime without violating fundamental physical requirements such as causality and thermodynamic stability. Consequently, the pressure at high densities must be compatible with the pQCD calculations of Gorda \textit{et al.}~\cite{Gorda:2018gpy,Gorda:2021kme,Gorda:2023mkk}, while the speed of sound remains subluminal. Together, these observational and theoretical constraints substantially restrict the class of viable EoSs.

Several previous studies have investigated hybrid-star models based on the extended linear sigma model (eLSM), demonstrating that they provide a realistic description of dense quark matter while remaining consistent with both lattice-QCD results at low baryon chemical potential and perturbative QCD constraints at high densities~\cite{Takatsy:2019vjs,Kovacs:2021ger,Takatsy:2023xzf,Kasza:2025lsc,Kasza:2026peb}. In the present work we considerably extend these investigations by systematically combining the eLSM quark matter description with several qualitatively different hadronic EoSs, including relativistic mean-field, chiral mean-field and meta-model based descriptions. The resulting hybrid EoSs are subjected to theoretical consistency requirements and current astrophysical constraints before their parameter space is explored using Bayesian inference.

The primary objective of this work is to determine which combinations of hadronic and quark matter remain compatible with present observations, identify the preferred hadron--quark crossover parameters, and investigate whether current astrophysical data already provide evidence for the formation of pure quark cores inside NSs. We further examine the implications of the resulting maximum-mass distributions for the nature of the secondary compact object in GW190814~\cite{LIGOScientific:2020zkf,Biswas:2020xna,Mali:2025etk}.

\section{Astrophysical constraints from NICER and GW170817}

Recent advances in X-ray and gravitational-wave astronomy have significantly improved our knowledge of cold dense matter. In the present work, we employ both electromagnetic and gravitational-wave observations to constrain hybrid NS EoSs.

The NS Interior Composition Explorer (NICER) mission determines the masses and radii of rotation-powered millisecond pulsars through Bayesian pulse-profile modelling of their thermal X-ray emission. The observed pulse profiles are shaped by general relativistic light bending and therefore depend sensitively on the stellar compactness. By comparing synthetic pulse profiles with the observed X-ray data, NICER infers the posterior probability distribution of the stellar mass and radius, providing direct constraints on the NS mass--radius relation predicted by a given EoS.

In this work, we employ recent mass--radius constraints based on NICER observations of five rotation-powered millisecond pulsars:
PSR~J0030+0451~\cite{Vinciguerra:2023qxq},
PSR~J0740+6620~\cite{Salmi:2024aum},
PSR~J0614$-$3329~\cite{Mauviard:2025dmd},
PSR~J0437$-$4715~\cite{Choudhury:2024xbk},
and PSR~J1231$-$1411~\cite{Qi:2025mpn}.

We note that a recent reanalysis of PSR~J0437$-$4715 by
Miller \textit{et al.}~\cite{Miller:2025qfq}, which explicitly models a modulated nonthermal emission component that was not considered in the analysis used in the present work, inferred a considerably broader radius posterior extending toward larger radii. 

We also note that a recent analysis of PSR~J0614$-$3329 by Miller \textit{et al.}~\cite{Miller:2026vpr}, currently available as an arXiv preprint, obtained a broader radius posterior extending toward larger radii than the one adopted in the present work. A quantitative assessment of the impact of this result, together with the updated posterior for PSR~J0437$-$4715, on our Bayesian inference is left for future work.

For PSR~J1231$-$1411, we adopt the recently published independent reanalysis by Qi \textit{et al.}~\cite{Qi:2025mpn}
for our fiducial analysis, while the sensitivity to this choice is discussed below. The original NICER analysis of PSR~J1231$-$1411~\cite{Salmi:2024bss} inferred a comparatively large stellar radius, making this source difficult to reconcile with the remaining NICER measurements using a single EoS. However, the independent reanalysis of the same NICER and XMM-Newton observations by Qi \textit{et al.}~\cite{Qi:2025mpn} inferred a substantially smaller stellar radius,
\begin{equation}
M = 1.12 \pm 0.07\,M_\odot,\qquad
R_{\rm eq}=9.91^{+0.88}_{-0.86}\ {\rm km},
\end{equation}
corresponding to the 68\% credible interval \cite{Qi:2025mpn}. The
updated posterior is considerably more consistent with the remaining NICER measurements and allows all currently available NICER mass--radius observations to be simultaneously described by a single
EoS. For this reason, we employ the updated posterior distribution of Ref.~\cite{Qi:2025mpn} in our fiducial analysis.
Nevertheless, given the substantial difference between the two J1231$-$1411 analyses and the complexity of this source discussed in Ref.~\cite{Salmi:2024bss}, we assess the sensitivity of our results to this choice by performing additional calculations using the alternative posterior of Ref.~\cite{Salmi:2024bss} and excluding PSR J1231$-$1411
altogether. The impact of these alternative treatments on our main results is briefly discussed in Sections~\ref{sec:4} and~\ref{sec:5}.

\begin{figure}
    \centering
    \includegraphics[scale=0.6]{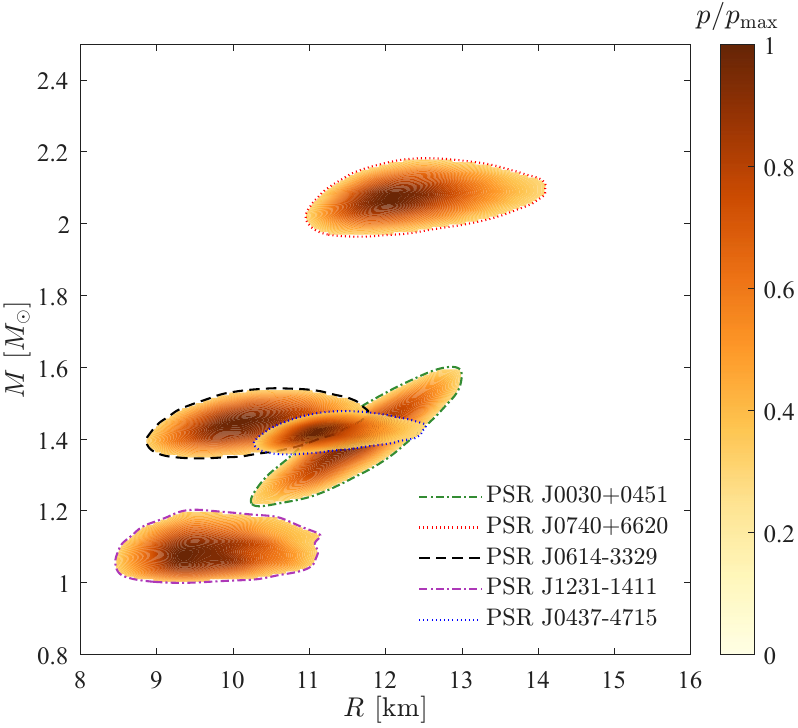}
    \includegraphics[scale=0.6]{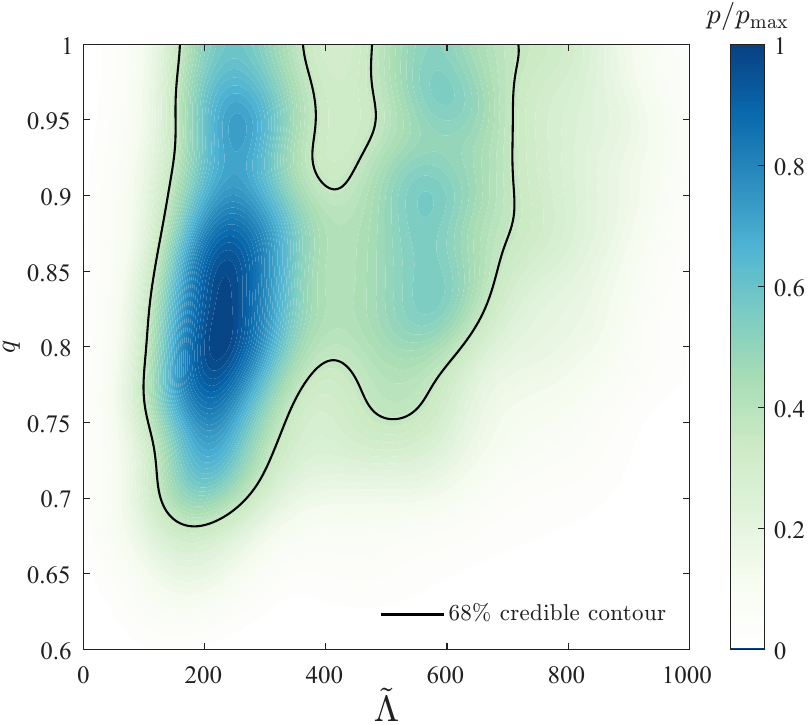}    
    \caption{\textit{Upper panel:} The NICER posterior distributions for the five pulsars. Only the 68\% credible contours are shown to reduce overlap between the posterior distributions and improve visual clarity.~\cite{Qi:2025mpn,Salmi:2024aum,Vinciguerra:2023qxq,Mauviard:2025dmd,Choudhury:2024xbk} \textit{Lower panel:} The LVC posterior distribution for GW170817 in the $\tilde{\Lambda}$--$q$ plane, obtained under the low-spin prior ($\chi<0.05$). The 68\% credible contour is indicated by the black line.~\cite{LIGOScientific:2018hze} In both panels, the posterior distributions are normalized to their respective maximum values, $p/p_{\mathrm{max}}$.}
    \label{fig:MRGWposteriors}
\end{figure}

The resulting mass--radius posterior distributions used in the present work are shown in Fig.~\ref{fig:MRGWposteriors}. It is important to emphasize that our Bayesian analysis utilizes the full posterior probability distributions, rather than only their central values and credible intervals, thereby preserving the complete observational information.

Gravitational-wave observations provide an independent and complementary probe of the NS EoS. During the inspiral phase of a binary NS merger, each star is tidally deformed by the gravitational field of its companion. These tidal deformations modify the orbital evolution and consequently leave measurable imprints on the emitted gravitational-wave signal.

The leading tidal contribution to the gravitational wave phase evolution is described by the binary effective tidal deformability,
\begin{equation}
\tilde{\Lambda} = \frac{16}{13} \frac{(M_1+12M_2)M_1^4\Lambda_1 + (M_2+12M_1)M_2^4\Lambda_2}{(M_1+M_2)^5},
\end{equation}
where $M_i$ and $\Lambda_i$ denote the masses and dimensionless tidal deformabilities of the two NSs, respectively~\cite{LIGOScientific:2018hze}. The binary mass
ratio is defined as
\begin{equation}
q=\frac{M_2}{M_1},
\qquad (M_2 \le M_1).
\end{equation}

The event GW170817 provided the first observational constraint on the binary effective tidal deformability and therefore on the high-density EoS of NS matter. In the present work we employ the posterior probability distribution in the
$(q,\tilde{\Lambda})$ plane published by the LIGO--Virgo Collaboration~\cite{LIGOScientific:2018hze}. Rather than applying only a single upper limit on
$\tilde{\Lambda}$, the complete posterior distribution is incorporated into our Bayesian likelihood, allowing the gravitational-wave information to be combined consistently with the NICER mass--radius measurements.

The NICER and GW170817 posterior distributions employed throughout in this work are presented in Fig.~\ref{fig:MRGWposteriors}. Their incorporation into the Bayesian framework is described in Sec.~\ref{sec:4}.

\section{Hybrid equation of state}

Recent years have witnessed significant progress in the development of unified hybrid equations of state that consistently describe hadronic and quark degrees of freedom within a single theoretical framework~\cite{Marczenko:2020jma,Clevinger:2022xzl,Yang:2023duo}. In contrast, the present work follows a phenomenological crossover approach to investigate the possible presence of deconfined quark matter in NS cores. We construct a family of hybrid EoSs by combining independent descriptions of hadronic and quark matter through a smooth hadron--quark crossover, allowing us to explore a broad range of transition densities and crossover widths while preserving thermodynamic consistency.

In this work, the quark phase is described within the extended Linear Sigma Model (eLSM), which provides an effective realization of low-energy QCD incorporating the approximate chiral symmetry of the strong interaction and its spontaneous breaking. In addition to scalar and pseudoscalar mesons, the model includes vector mesons and constituent quark degrees of freedom, allowing a unified description of the restoration of chiral symmetry at high baryon densities.
The thermodynamic quantities of quark matter are obtained within the mean-field approximation. A detailed description of the model can be found in Ref.~\cite{Parganlija:2012fy}.

For the hadronic phase we employ four representative EoSs that cover qualitatively different microscopic descriptions of dense nuclear matter. These EoSs are discussed in detail in Sec.~\ref{sec:3B}.

Throughout this work, both the hadronic and quark-matter EoSs are constructed under the conditions of beta equilibrium and electric charge neutrality, as appropriate for cold neutron-star matter.

The following subsections summarize the quark and hadronic models before introducing the crossover construction used throughout the present work.

\subsection{Quark matter}

The quark phase is described by the extended Linear Sigma Model (eLSM)~\cite{Parganlija:2012fy} with constituent quarks, which has been successfully applied to the study of strongly interacting matter over a wide range of temperatures and baryon densities. The model realizes chiral symmetry linearly and incorporates its spontaneous breaking through nonvanishing scalar condensates, while successfully reproducing lattice QCD results at zero chemical potential and remaining compatible with available lattice calculations for baryon chemical potentials up to $\mu_{\rm B} \le 400\,\text{MeV}$~\cite{Kovacs:2016juc,Kovacs:2017jjg}.

An important ingredient of the present implementation is the repulsive vector interaction between quarks, governed by the vector coupling strength $g_V$. This interaction stiffens the quark EoS at high densities and therefore has a significant influence on the maximum mass and radius of hybrid NSs. Since $g_V$ does not enter the hadron masses and decay widths used to determine the remaining model parameters, it is not constrained by the vacuum fit and is therefore treated as a free parameter in the Bayesian analysis.

Within the mean-field approximation, the grand thermodynamic potential is minimized with respect to the mesonic condensates, yielding the pressure, energy density, baryon density, and other thermodynamic quantities required for the construction of NS EoSs. The resulting quark matter EoS satisfies thermodynamic consistency and provides the high-density component of the hybrid models investigated in this work.

Although the eLSM contains a relatively large number of model parameters, most of them are tightly constrained by low- and intermediate-energy hadronic observables and therefore remain fixed throughout the present work. The scalar-isoscalar $\sigma$-meson mass is fixed at $m_\sigma = 290\,\mathrm{MeV}$, as this value was favored by \cite{Kovacs:2016juc} and a previous Bayesian analysis of NS observations~\cite{Takatsy:2023xzf}. Consequently, the vector coupling strength $g_V$ is the only quark-sector parameter treated as free in the Bayesian analysis.

\subsection{Hadronic matter}\label{sec:3B}

To account for the remaining theoretical uncertainties in the hadronic
EoS, we consider four representative models spanning
qualitatively different microscopic descriptions of dense nuclear matter. Rather than comparing different parameterizations of the same theoretical framework, we deliberately select EoSs based on distinct physical assumptions. This strategy enables us to assess the robustness of our conclusions with respect to the modeling of the hadronic phase.

The hadronic EoSs employed in this work are summarized in Table~\ref{tab:hadronicEOS}.

\begin{table*}
\centering
\caption{Summary of the hadronic EoSs employed in this work.}
\label{tab:hadronicEOS}
\begin{ruledtabular}
\begin{tabular}{lllll}
Model &
Framework &
Degrees of freedom &
Crust &
Characteristics\\
\hline

SFHo~\cite{Hempel:2009mc,Steiner:2012rk} &
RMF &
Nucleons &
SLy4~\cite{Douchin:2001sv} &
Soft EoS \\

DD2~\cite{Hempel:2009mc,Typel:2009sy} &
Density dependent RMF &
Nucleons &
SLy4~\cite{Douchin:2001sv} &
Stiff EoS \\

DNS(CMF)~\cite{Dexheimer:2008ax, Schurhoff:2010ph} &
Chiral mean field &
Nucleons + Hyperons &
Unified &
Hyperonic EoS \\

GMSR(BSK16)~\cite{Grams:2022lci} &
Nuclear meta-model + CLDM &
Nucleons &
Unified &
BSK16-inspired functional \\

\end{tabular}
\end{ruledtabular}
\end{table*}

The SFHo EoS is based on the relativistic mean-field (RMF) framework,with nucleons interacting through the exchange of $\sigma$, $\omega$, and $\rho$ mesons, including nonlinear self-interaction terms~\cite{Hempel:2009mc,Steiner:2012rk}. Its parameters were calibrated to reproduce empirical nuclear-matter properties, while also accounting for astrophysical constraints available at the time of its development. For the crust, the SFHo EoS is matched to the SLy4 EoS~\cite{Douchin:2001sv}. The resulting nuclear-matter properties lead to a comparatively soft hadronic EoS, which is reflected in the relatively small NS radii predicted by SFHo.

The DD2 model is also formulated within the RMF approach but employs density-dependent meson--nucleon couplings~\cite{Typel:2009sy}. In the present work, we use the homogeneous nucleonic DD2 EoS, matched to the SLy4 EoS~\cite{Douchin:2001sv} in the crust. Thus, the light-cluster degrees of freedom considered in Ref.~\cite{Typel:2009sy} are not included in the DD2 sector employed here. Compared with SFHo, DD2 predicts systematically larger NS radii and maximum masses, providing a representative example of a relatively stiff hadronic EoS.

DNS(CMF) is based on the SU(3) Chiral Mean Field (CMF) model, in which the effective hadron masses originate from spontaneous chiral symmetry breaking~\cite{Dexheimer:2008ax, Schurhoff:2010ph}. The model naturally incorporates hyperons at high baryon densities while maintaining thermodynamic consistency over the entire density range. Consequently, DNS(CMF) enables us to investigate the impact of hyperonic degrees of freedom on the properties of hybrid NSs.

The GMSR(BSK16) EoS is constructed within a nuclear meta-modeling framework calibrated to reproduce the microscopic BSK16 energy-density functional~\cite{Grams:2022lci}. The description of the NS crust is provided by the Compressible Liquid Drop Model (CLDM), resulting in a unified EoS covering both the crust and the core. In contrast to the other hadronic models considered here, GMSR(BSK16) is not derived from a relativistic mean-field theory, thereby providing an independent benchmark for assessing the sensitivity of our results to the underlying nuclear physics.

Together, these four hadronic models span a broad range of theoretical
approaches, from relativistic mean-field and chiral mean-field theories to nuclear meta-modeling. Their diversity allows us to quantify the dependence of the inferred hadron--quark crossover parameters on the uncertainties of the hadronic EoS.

\subsection{Hadron--quark crossover}

The hadronic and quark EoSs are connected through a smooth hadron--quark crossover rather than a first-order phase transition. Such a construction avoids the appearance of a mixed phase with constant pressure and instead provides a continuous interpolation between hadronic and quark matter over a finite density interval. The crossover is characterized by two independent parameters: the central transition density, $\bar{\rho}$, and the transition half-width, $\Gamma$.

Following our previous work, the interpolation is performed for the energy density rather than directly interpolating the pressure or other thermodynamic quantities. This choice guarantees a thermodynamically consistent construction of the hybrid EoS. The energy density within the crossover region is represented by a fifth-order polynomial,
\begin{equation}
\varepsilon(\rho)=\sum_{i=0}^{5}a_i\rho^i,
\end{equation}
where $\rho$ denotes the baryon density. A fifth-order polynomial is the lowest-order function that simultaneously satisfies the continuity conditions imposed on the energy density, pressure, and sound speed at both boundaries of the crossover region. The polynomial coefficients are uniquely determined by requiring the continuity of these thermodynamic quantities at the lower and upper boundaries,
\begin{equation}
\rho_{\rm low}=\bar{\rho}-\Gamma,
\qquad
\rho_{\rm high}=\bar{\rho}+\Gamma.
\end{equation}

Below the transition region the EoS is given by the selected hadronic model, whereas above the transition region the thermodynamic
quantities are obtained from the eLSM. The polynomial interpolation provides a smooth connection between the two limiting descriptions while preserving thermodynamic consistency throughout the entire density range.

Once the energy density has been constructed, the pressure is obtained from the thermodynamic relation
\begin{equation}
P(\rho)=\rho\frac{\partial\varepsilon}{\partial\rho}-\varepsilon,
\end{equation}
from which all remaining thermodynamic quantities entering the Tolman--Oppenheimer--Volkoff equations can be calculated. In particular, the
adiabatic sound speed is evaluated as
\begin{equation}
c_s^2=\frac{\partial P}{\partial\varepsilon},
\end{equation}
allowing us to verify both thermodynamic stability and causality for every hybrid EoS considered in this work.

The crossover construction therefore introduces three free parameters into the hybrid EoS: the vector coupling strength of the quark
model, $g_V$, the central transition density, $\bar{\rho}$, and the crossover width, $\Gamma$. These parameters define the model space explored
in the Bayesian analysis presented in the following sections.

\section{Bayesian Framework and EoS Selection}\label{sec:4}

In this section, we first introduce the Bayesian framework used to constrain the hybrid-EoS parameters. We then describe the theoretical and observational constraints applied to the parameter space and, finally, compare the four hadronic EoSs and select the model used for the detailed Bayesian analysis.

\subsection{Bayesian methodology}\label{sec:4a}

Within the Bayesian framework adopted in this work, the hybrid-EoS parameterizations satisfying the theoretical prior are weighted using astrophysical observations.

The posterior probability of a parameter set $\theta$ is obtained from Bayes' theorem,
\begin{equation}
P(\theta|D)=
\frac{P(D|\theta)\,P(\theta)}
{\int P(D|\theta)\,P(\theta)\,d\theta},
\end{equation}
where $D$ denotes the complete set of astrophysical observations, $P(\theta)$ is the prior probability defined in the previous subsection, and $P(D|\theta)$ is the likelihood of obtaining the observational data for the parameter set $\theta$.\footnote{An alternative Bayesian framework has recently been introduced in which the prior is defined directly in the NS mass--radius plane rather than in the EoS parameter space. The corresponding posterior is subsequently mapped back to the EoS using accurate inversion techniques, providing a more natural prior and reducing sensitivity to the assumed EoS parameterization.~\cite{Sun:2026itt}}

Since only relative posterior probabilities are required throughout this work, the normalization constant in the denominator is omitted, yielding
\begin{equation}
P(\theta|D)\propto P(D|\theta)\,P(\theta).
\end{equation}

Assuming statistical independence of the individual measurements, the total likelihood is written as
\begin{equation}
P(D|\theta)
=
P_{\rm GW}(\theta)
\prod_i
P_{{\rm NICER},i}(\theta),
\end{equation}
where the product extends over all NICER observations considered in this
work.

For each NICER pulsar, the likelihood is evaluated by sampling the published two-dimensional posterior distribution in the mass--radius plane along the theoretical mass--radius relation predicted by the given EoS,
\begin{equation}
P_{\rm NICER}(\theta)
\propto
\int
P_{\rm N}(M,R_\theta(M))
\,dM,
\end{equation}
where $P_{\rm N}(M,R)$ is the published NICER posterior probability density in the mass--radius plane, $M$ is the gravitational mass of the NS, and $R_\theta(M)$ is the corresponding theoretical radius predicted by the parameter set $\theta$.

The likelihood associated with GW170817 is calculated from the published joint posterior distribution of the effective tidal deformability $\tilde{\lambda}$ and the binary mass ratio $q$,
\begin{equation}
P_{\rm GW}(\theta)
\propto
\int_{M_{\rm eq}}^{M_{\rm TOV,\theta}}
dM_1\,
P_{\rm LVC}
\left[
\tilde{\Lambda}_{\theta}(M_1,\mathcal{M}),
q(M_1,\mathcal{M})
\right].
\end{equation}
Here, $P_{\rm LVC}(\tilde{\Lambda},q)$ denotes the published posterior probability density from GW170817, $M_1$ is the mass of the heavier binary component, while the mass of the second component is determined from $M_1$ and the measured chirp mass $\mathcal{M}$. The chirp mass is defined as
\begin{equation}
\mathcal{M}
=
\frac{(M_1M_2)^{3/5}}
{(M_1+M_2)^{1/5}},
\end{equation}
where $M_2\leq M_1$ is the mass of the lighter component. Furthermore, $\tilde{\Lambda}_{\theta}(M_1,\mathcal{M})$ is the effective tidal
deformability predicted by the EoS after determining $M_2$ from the fixed value of $\mathcal{M}$. The lower integration limit,
\begin{equation}
M_{\rm eq}=2^{1/5}\mathcal{M},
\end{equation}
corresponds to the equal-mass configuration, $M_1=M_2$, while the upper
limit $M_{\rm TOV,\theta}$ denotes the maximum gravitational mass supported by the EoS specified by the parameter set $\theta$. Since each parameterization corresponds to a distinct value of $\theta$, the resulting maximum mass is evaluated separately for each parameterization.

The resulting posterior probabilities are subsequently used to construct the weighted distributions of the equation-of-state parameters and the derived NS observables presented in the following section.

\subsection{Theoretical and observational constraints on the hybrid EoS}

The parameter space of the hybrid EoS was explored through systematic scans over the three free parameters,
\[
\theta=(g_V,\Gamma,\bar{\rho}).
\]
In the first stage of the analysis, we employed a coarse parameter grid for each of the four hadronic EoSs considered in this work. The parameter ranges were chosen as $g_V \in [0,8]$, $\Gamma \in [1,6]\,\rho_0$, and
$\bar{\rho} \in [2,7]\,\rho_0$, with grid spacings
$\Delta g_V=0.5$, $\Delta\Gamma=0.2\,\rho_0$, and
$\Delta\bar{\rho}=0.2\,\rho_0$, respectively. The resulting grid contains 11492 parameter combinations before applying any physical constraints.
In addition, we required the lower boundary of the crossover region to satisfy
$\rho_{\rm low}=\bar{\rho}-\Gamma \geq 1\,\rho_0$.
The primary purpose of this coarse scan was to determine which hadronic EoS, when combined with the eLSM quark-matter EoS through the crossover construction, is most compatible with a set of imposed hard-cut constraints, described below.

Each parameter set was subjected to a series of theoretical and observational consistency requirements. As a first step, all EoSs were required to satisfy the fundamental physical conditions of thermodynamic stability and causality. Parameterizations violating either of these requirements were discarded.

We assume that each hybrid EoS is applicable up to the central baryon chemical potential of the maximum-mass NS configuration it supports.
Beyond this point, every hybrid EoS is required to admit a stable and causal matching to the pQCD EoS at high densities. Following Refs.~\cite{Takatsy:2023xzf,Kasza:2025lsc}, we test whether such a connection exists between the hybrid EoS at the maximum-mass configuration and a reference point obtained from the pQCD calculation of Gorda et al.~\cite{Gorda:2018gpy,Gorda:2021kme,Gorda:2023mkk}.
For this pQCD point, we use
$\mu_{\rm QCD}=2.6$ GeV,
$\rho_{\rm QCD}=6.47~{\rm fm}^{-3}$, and
$P_{\rm QCD}=3823~{\rm MeV\,fm}^{-3}$.

Specifically, we first require
\begin{equation}
\rho_{\rm c,TOV} \leq \rho_{\rm QCD}, \qquad
P_{\rm c,TOV} \leq P_{\rm QCD},
\end{equation}
which ensures that the baryon density and pressure increase between the two endpoints, together with
\begin{equation}
\frac{\rho_{\rm c,TOV}}{\mu_{\rm c,TOV}}
\leq
\frac{\rho_{\rm QCD}}{\mu_{\rm QCD}},
\end{equation}
which follows from the requirement of causality along the connecting EoS. Here $(\mu_{\rm c,TOV},P_{\rm c,TOV},\rho_{\rm c,TOV})$ denote the baryon chemical potential, pressure and baryon density at the center of the maximum-mass NS configuration. In addition, the total pressure difference between the two endpoints, $\Delta P=P_{\rm QCD}-P_{\rm c,TOV}$, must satisfy the integral constraint
\begin{equation}
\Delta P_{\rm min} \leq \Delta P \leq \Delta P_{\rm max},
\end{equation}
where
\begin{align}
\Delta P_{\rm min}
&=
\frac{\mu_{\rm QCD}^{2}-\mu_{\rm c,TOV}^{2}}{2}
\frac{\rho_{\rm c,TOV}}{\mu_{\rm c,TOV}},
\\
\Delta P_{\rm max}
&=
\frac{\mu_{\rm QCD}^{2}-\mu_{\rm c,TOV}^{2}}{2}
\frac{\rho_{\rm QCD}}{\mu_{\rm QCD}}.
\end{align}
This integral constraint ensures that the total pressure increase between the two endpoints is compatible with a causal and thermodynamically stable EoS connecting them. Parameterizations for which these conditions are not satisfied are excluded from further consideration.

These theoretical requirements (stability, causality and pQCD-consistency) define the prior used throughout the Bayesian analysis. In other words, parameterizations satisfying the stability, causality, and pQCD consistency conditions were assigned a uniform prior, whereas all remaining parameterizations were assigned zero probability.

Depending on the physical quantity under investigation, additional astrophysical constraints were imposed before constructing the posterior distributions.
\begin{itemize}
\item a lower bound on the maximum NS gravitational mass,
      $M_{\rm TOV}>2.004\,M_\odot$,
      corresponding to the $1\sigma$ lower limit of the NICER mass
      measurement of PSR J0740+6620~\cite{Salmi:2024aum};

\item a more restrictive lower bound,
      $M_{\rm TOV}>2.27\,M_\odot$,
      motivated by recent Black Widow population studies~\cite{Romani:2025ytn};

\item an upper bound on the effective tidal deformability,
      $\tilde{\Lambda}<720$,
      obtained from the EOS-independent LIGO--Virgo analysis of GW170817~\cite{LIGOScientific:2018hze}.
\end{itemize}

These constraints were applied selectively according to the quantity being studied. For example, when investigating the posterior distribution of the maximum NS mass, no constraint on $M_{\rm TOV}$ was imposed in order to avoid circular reasoning.

\subsection{Comparison and selection of the hadronic EoS}\label{sec:4c}

We first compare the four hadronic EoSs in order to assess how the choice of the hadronic sector affects the compatibility of the resulting hybrid EoSs with the astrophysical constraints and to identify which of them provides the best agreement with the astrophysical observations considered in this work.

Table~\ref{tab:hardcuts} summarizes the number of parameterizations
surviving the successive hard cuts for each of the four hadronic EoSs on the coarse grid. The results reveal a pronounced dependence on the underlying hadronic EoS. Neither DD2 nor DNS(CMF) admits a parameterization that simultaneously satisfies the more restrictive
$M_{\rm TOV}>2.27\,M_\odot$ and $\tilde{\Lambda}<720$ constraints. Even when the weaker mass constraint,
$M_{\rm TOV}>2.004\,M_\odot$, is adopted, no viable DNS(CMF) parameterization remains, while only 19 parameterizations based on the DD2 EoS satisfy both constraints. In contrast, both SFHo and GMSR(BSK16) retain a substantial number of parameterizations even when the more restrictive mass constraint is combined with the tidal-deformability bound.

\begin{table*}[t]
\caption{
Number of surviving parameterizations on the coarse grid after the successive application of the theoretical prior and observational hard constraints for the four hadronic EoSs.
}
\label{tab:hardcuts}
\begin{ruledtabular}
\begin{tabular}{lcccc}
Selection step & SFHo & GMSR(BSK16) & DD2 & DNS(CMF) \\
\hline
Prior
    & 2775 & 1946 & 2800 & 3027 \\
Prior + $\tilde{\Lambda}<720$
    & 2471 & 1782 & 535 & 231 \\
Prior + $\tilde{\Lambda}<720$ +$M_{\rm TOV}>2.004\,M_\odot$ & 711 & 636 & 19 & 0 \\
Prior + $\tilde{\Lambda}<720$ + $M_{\rm TOV}>2.27\,M_\odot$
    & 234 & 274 & 0 & 0 \\
\end{tabular}
\end{ruledtabular}
\end{table*}

\begin{figure}[h!]
    \centering
    \includegraphics[scale=0.6]{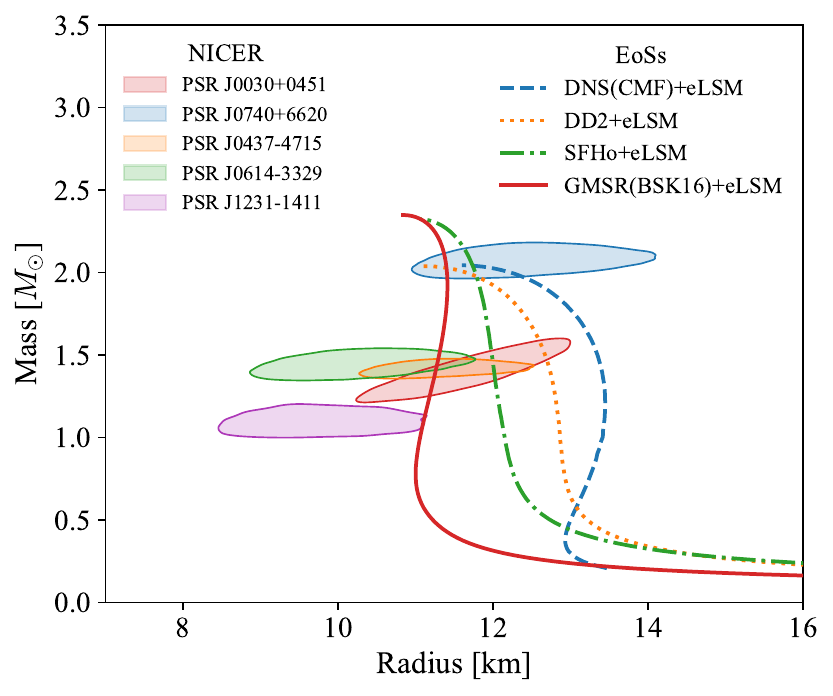}
    \caption{Mass--radius relations corresponding to the highest-probability crossover parameterization obtained for each of the four hadronic EoSs on the coarse parameter grid. The best parameterizations were selected from the prior set using the joint likelihood of the five NICER measurements and GW170817. The colored areas indicate the 1$\sigma$ credible regions of the five NICER analyses considered in this work.}
    \label{fig:best_eos_param}
\end{figure}

To further quantify the relative agreement with the astrophysical data, we determined the highest-probability parameterization within the prior set for each hadronic EoS. The joint likelihood was constructed using the five NICER measurements considered in this work together with the GW170817 gravitational-wave observation, as described in Sec.~\ref{sec:4a}. Table~\ref{tab:eos_prob} lists the probability of the best crossover parameterization obtained for each hadronic EoS, normalized to the largest probability among all crossover parameterizations constructed with the four hadronic EoSs and belonging to the prior set. The global maximum is obtained for GMSR(BSK16). The best SFHo parameterization has a relative probability of approximately $3.8\times10^{-2}$, whereas the corresponding values for DD2 and DNS(CMF) are much lower. The mass--radius relations corresponding to these four parameterizations are shown in Fig.~\ref{fig:best_eos_param}.

\begin{table}[h!]
\caption{Relative probability of the highest-probability parameterization obtained for each hadronic EoS on the coarse grid. Here $\mathcal{P}_{\rm max}$ denotes the largest joint probability within the prior set across all four hadronic EoSs.}
\label{tab:eos_prob}
\begin{ruledtabular}
\begin{tabular}{lc}
Hadronic EoS & $p/\mathcal{P}_{\rm max}$ \\
\hline
GMSR(BSK16) & $1$ \\
SFHo        & $3.80\times10^{-2}$ \\
DD2         & $1.99\times10^{-4}$ \\
DNS(CMF)    & $1.86\times10^{-7}$ \\
\end{tabular}
\end{ruledtabular}
\end{table}

The behavior of the preferred crossover parameters provides additional insight into the strong dependence on the hadronic baseline. For both DNS(CMF) and DD2, the highest-probability configurations are driven toward a very early and extremely broad crossover:
$\rho_{\rm low}=\bar{\rho}-\Gamma$ approaches its lower limit of approximately $1\,\rho_0$, whereas
$\rho_{\rm high}=\bar{\rho}+\Gamma$ is driven toward the upper boundary of the explored range. We verified this behavior by successively extending the parameter ranges; the preferred $\rho_{\rm high}$ follows the extended upper boundary while the probability continues to increase.
This indicates that, for these hadronic EoSs, the astrophysical data do not localize a finite crossover interval within the explored physically reasonable domain. Instead, the crossover is driven toward an increasingly early onset and broad transition in an attempt to compensate for the less favorable properties of the underlying hadronic EoS.
Nevertheless, even the best DD2 and DNS(CMF) configurations remain substantially less probable than their GMSR(BSK16) and SFHo counterparts.
This behavior is also apparent from the mass--radius relations in Fig.~\ref{fig:best_eos_param}: DD2 and DNS(CMF) generally predict larger radii and consequently show substantially less overlap with the 1$\sigma$ regions preferred by the NICER measurements.

Importantly, the strong suppression of the DD2- and DNS(CMF)-based parameterizations is not caused by the adopted J1231$-$1411 posterior. As shown by the successive hard-cut analysis, these EoSs already have difficulty simultaneously satisfying the tidal-deformability and
maximum-mass constraints, independently of the NICER likelihood. Alternative analyses of J1231$-$1411~\cite{Salmi:2024bss} and
J0437$-$4715~\cite{Miller:2025qfq}, both of which favor larger radii than the posteriors adopted in our fiducial analysis, could provide a more favorable NICER likelihood for these relatively stiff hadronic EoSs. This would not, however, affect their independent tension with the $\tilde{\Lambda}<720$ and maximum-mass constraints.

In contrast, for the softer SFHo and GMSR(BSK16) hadronic EoSs, the preferred crossover parameters are localized at intermediate values within the explored parameter ranges. These models also provide substantially larger joint probabilities. The comparison therefore indicates a strong dependence of the inferred crossover properties on the stiffness of the underlying hadronic EoS. In particular, the astrophysical data favor a finite crossover region for SFHo and GMSR(BSK16), whereas for DD2 and DNS(CMF) even an increasingly early and broad crossover does not provide a comparably good description of the data, as clearly demonstrated by the relative probabilities reported in Table~\ref{tab:eos_prob}. It should be noted, however, that the SFHo-based model exhibits some tension with the specific NICER posteriors adopted in our fiducial analysis, particularly because of its difficulty in reproducing radii below approximately 12 km for NSs around $1.4\,M_\odot$. This conclusion is sensitive to the choice of observational posteriors. In particular, the alternative analyses of PSR J1231$-$1411 by Salmi et al.~\cite{Salmi:2024bss} and PSR J0614$-$3329 by Miller et al.~\cite{Miller:2026vpr} allow larger stellar radii and may therefore reduce the tension with
SFHo-based constructions.

However, it should be emphasized that in each case we considered a single publicly available parameterization of the hadronic model taken from the CompOSE database. Consequently, the present analysis does not establish whether the tension with observations originates from the underlying theoretical framework itself or merely from the specific parameterization employed. For example, the DNS(CMF) parameterization adopted here assumes a nuclear incompressibility of $K=300$ MeV, which lies above the range generally preferred by low-energy nuclear experiments. Such a large incompressibility leads to a comparatively stiff hadronic EoS and may facilitate the support of high maximum NS masses. Within the framework of hybrid EoS, however, similarly large maximum masses can be achieved without requiring such a stiff hadronic sector, owing to the additional high-density contribution from deconfined quark matter.

Based on the coarse-grid comparison, GMSR(BSK16) was selected for the detailed Bayesian analysis. To ensure that the inferred posterior distributions are sufficiently well resolved, we repeated the scan for the eLSM+GMSR(BSK16) model using a denser and extended three-dimensional grid. We adopted
$g_V\in[0,8]$, $\Gamma\in[1,9]\,\rho_0$, and
$\bar{\rho}\in[2,10]\,\rho_0$, with spacings
$\Delta g_V=0.25$, $\Delta\Gamma=0.2\,\rho_0$, and
$\Delta\bar{\rho}=0.1\,\rho_0$. This corresponds to 109593 parameter combinations before imposing the condition $\bar{\rho}-\Gamma\geq1\,\rho_0$.
The extended and refined grid serves three purposes:
(i) to increase the number of theoretical EoSs entering the Bayesian analysis, (ii) to resolve possible narrow or strongly correlated structures in the posterior distribution, and (iii) to ensure that the marginalized posterior distributions are sufficiently contained within the adopted parameter ranges for reliable determination of credible intervals.

After imposing the theoretical prior, 19589 GMSR(BSK16)-based parameterizations remain on this refined grid. As for the coarse-grid analysis, the observational hard cuts were applied successively, and the numbers of EoS parameterizations remaining after each selection step are summarized in Table~\ref{tab:refined_cuts}. The maximum of the joint posterior is located at
\[
(g_V,\Gamma,\bar{\rho})
=
(5.25,\,3.8\,\rho_0,\,6.1\,\rho_0).
\]
For comparison, the corresponding maximum obtained from the coarse scan was
\[
(g_V,\Gamma,\bar{\rho})
=
(5.0,\,3.6\,\rho_0,\,5.8\,\rho_0).
\]
The shift of the maximum under grid refinement motivated an additional convergence test.

For this purpose, we performed a further local scan covering approximately the $1\sigma$ region around the preferred parameter range, reducing the grid spacings to
$\Delta g_V=0.125$,
$\Delta\Gamma=0.1\,\rho_0$, and
$\Delta\bar{\rho}=0.05\,\rho_0$.
This high-resolution scan was used solely as a convergence test and not to redefine the parameter domain of the Bayesian analysis. The resulting maximum was found at
\[
(g_V,\Gamma,\bar{\rho})
=
(5.325,\,3.8\,\rho_0,\,6.1\,\rho_0).
\]
Thus, $\Gamma$ and $\bar{\rho}$ remain unchanged, while the shift in $g_V$ is smaller than the grid spacing of the refined scan. We therefore find that the location of the maximum-posterior region is stable under further grid refinement.

We additionally verified that the preferred parameterization on the extended grid is insensitive to the subsequent application of the observational hard cuts. Starting from the prior set, imposing $\tilde{\Lambda}<720$ together with either of the two maximum-mass constraints leaves the maximum-posterior parameterization unchanged at
$(g_V,\Gamma,\bar{\rho})=(5.25,3.8\,\rho_0,6.1\,\rho_0)$.

Based on these results, the detailed Bayesian inference presented in the remainder of this work is performed on the extended and refined eLSM+GMSR(BSK16) parameter grid.

\begin{table}[t]
\caption{
Number of GMSR(BSK16)-based EoS parameterizations remaining on the refined parameter grid after the successive application of the theoretical prior and observational hard cuts.
}
\label{tab:refined_cuts}
\begin{ruledtabular}
\begin{tabular}{lc}
Selection step & \# \\
\hline
Prior
    & 19589 \\
Prior + $\tilde{\Lambda}<720$
    & 19052 \\
Prior + $\tilde{\Lambda}<720$ + $M_{\rm TOV}>2.004\,M_\odot$
    & 4667 \\
Prior + $\tilde{\Lambda}<720$ + $M_{\rm TOV}>2.27\,M_\odot$
    & 1258  \\
\end{tabular}
\end{ruledtabular}
\end{table}

\section{Bayesian inference results}\label{sec:5}

Unless stated otherwise, all results presented in this section refer to nonrotating NS configurations obtained by solving the Tolman--Oppenheimer--Volkoff equations.

The theoretical and observational constraints introduced in the previous section substantially reduce the viable parameter space of the hybrid EoS. In particular, the DD2 and DNS(CMF) hadronic models fail to simultaneously satisfy the tidal deformability and maximum-mass constraints, while the remaining SFHo parameterizations generally predict radii around canonical NS masses that slightly exceed those preferred by the current NICER measurements. Consequently, GMSR(BSK16) emerges as the most strongly favored hadronic EoS among the models considered, based on the combined astrophysical likelihood adopted in our fiducial analysis. The remainder of this section therefore focuses on the Bayesian results obtained for the eLSM+GMSR(BSK16) hybrid EoS.

\subsection{Bayesian constraints on the hybrid EoS}

\begin{figure*}
    \centering
    \includegraphics[scale=0.43]{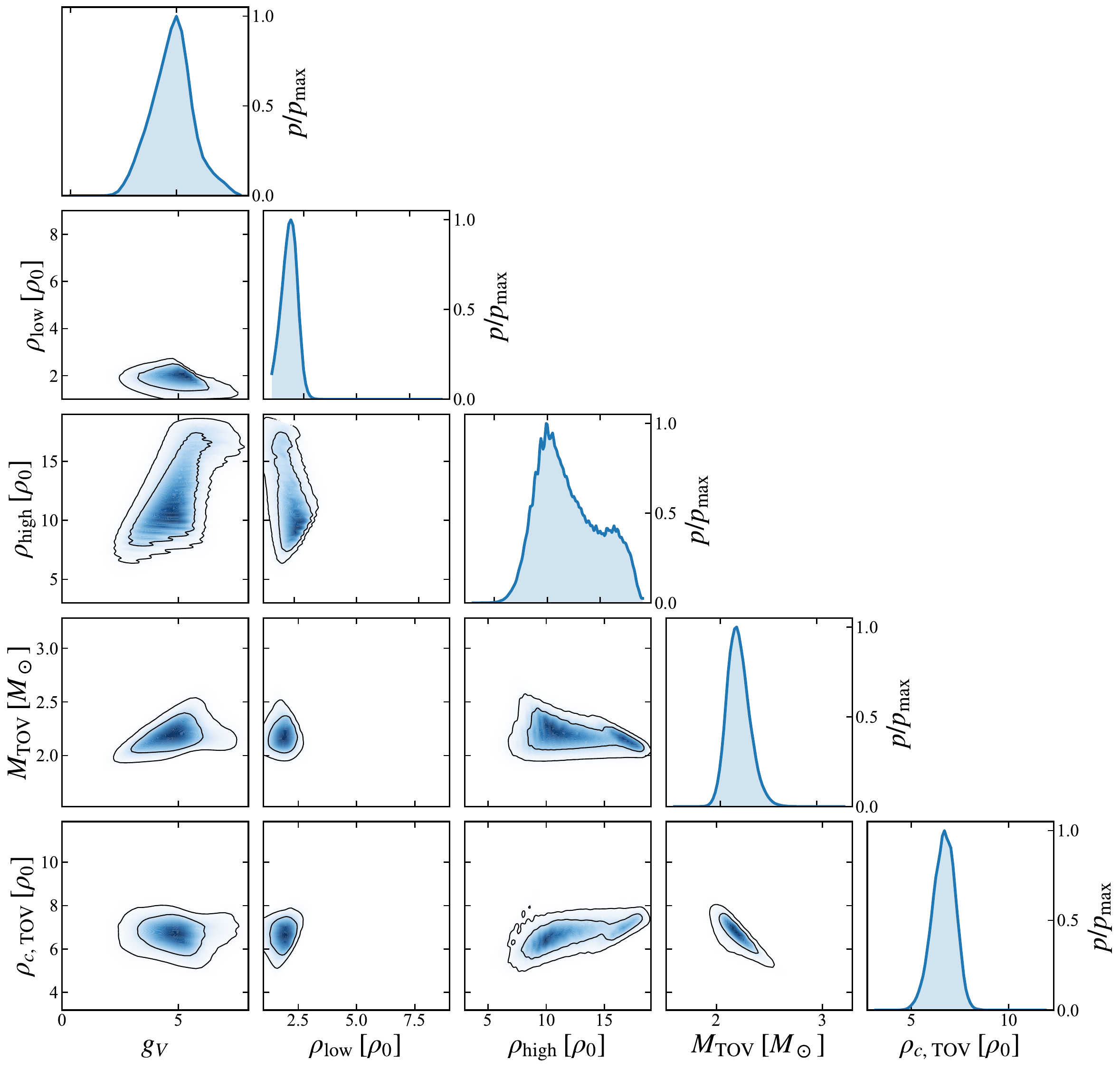}
    \caption{Marginalized posterior distributions for the
    eLSM+GMSR(BSK16) hybrid EoS obtained from the combined GW170817 and NICER likelihoods using the complete theoretical prior set, without additional observational hard cuts. The crossover region is parameterized by its lower and upper density boundaries, $\rho_{\rm low}=\bar{\rho}-\Gamma$ and
    $\rho_{\rm high}=\bar{\rho}+\Gamma$, respectively.
    The diagonal panels show the one-dimensional marginalized
    posterior distributions, while the off-diagonal panels show the corresponding two-dimensional distributions. The latter also include the 68\% and 95\% credible contours.}
    \label{fig:corner}
\end{figure*}

Figure~\ref{fig:corner} shows the marginalized posterior distributions obtained for the eLSM+GMSR(BSK16) hybrid EoS using the combined GW170817 and NICER likelihoods. Here we consider the complete theoretical prior set without imposing additional observational hard cuts on $\tilde{\Lambda}$ or $M_{\rm TOV}$, such that the distributions directly illustrate the constraints provided by the observational likelihoods on the model parameters and the resulting NS properties.

Instead of the original crossover parameters $\bar{\rho}$ and $\Gamma$, we characterize the transition region in terms of its lower and upper density boundaries, $\rho_{\rm low}$ and $\rho_{\rm high}$. This representation provides a more direct physical interpretation of the inferred crossover region, since $\rho_{\rm low}$ and $\rho_{\rm high}$ specify the densities at which the interpolation between the hadronic and quark-matter EoSs begins and ends, respectively. The posterior favors the onset of the crossover at relatively low densities, while allowing a substantially broader range for its upper boundary. The correlations visible in Fig.~\ref{fig:corner} therefore indicate that the observations constrain the onset of the crossover more strongly than its completion at high density. The figure also shows the corresponding posterior distributions of $M_{\rm TOV}$ and the central density
$\rho_{c,\rm TOV}$ of the maximum-mass configuration. The median values and corresponding 68\% credible intervals of the marginalized posterior distributions shown in Fig.~\ref{fig:corner}
are summarized in Table~\ref{tab:posterior_constraints}. We note that the maximum-posterior parameterization,
$(g_V,\Gamma,\bar{\rho})=(5.25,3.8\,\rho_0,6.1\,\rho_0)$, corresponds to $\rho_{\rm low}=2.3\,\rho_0$, which lies slightly outside the 68\% credible interval of the marginalized $\rho_{\rm low}$ distribution reported in Table~V. This does not indicate an inconsistency, since the former is derived from the maximum of the full multidimensional posterior, whereas the latter is obtained after marginalization over the remaining parameters. Correlations and posterior-volume effects can therefore shift the marginalized credible region relative to the value associated with the joint posterior maximum. The value $\rho_{\rm low}=2.3\,\rho_0$ remains within the corresponding 95\% credible interval.

\begin{table}
\caption{Median values and 68\% and 95\% credible intervals for the eLSM+GMSR(BSK16) hybrid EoS obtained from the marginalized posterior distributions shown in
Fig.~\ref{fig:corner}.}
\label{tab:posterior_constraints}
\centering
\normalsize
\begin{ruledtabular}
\begin{tabular}{lccc}
Quantity & Median & 68\% CI & 95\% CI \\
\hline
$g_V$
& $4.708$ & $[3.689,\,5.583]$ & $[2.717,\,6.856]$ \\

$\rho_{\rm low}\,[\rho_0]$
& $1.790$ & $[1.391,\,2.140]$ & $[1.033,\,2.469]$ \\

$\rho_{\rm high}\,[\rho_0]$
& $11.469$ & $[9.117,\,15.572]$ & $[7.365,\,17.838]$ \\

$M_{\rm TOV}\,[M_\odot]$
& $2.169$ & $[2.067,\,2.293]$ & $[1.974,\,2.441]$ \\

$\rho_{c,\rm TOV}\,[\rho_0]$
& $6.634$ & $[6.041,\,7.178]$ & $[5.401,\,7.654]$ \\
\end{tabular}
\end{ruledtabular}
\end{table}

\begin{figure*}[t]
    \centering

    \begin{minipage}[t]{0.49\textwidth}
        \centering

        \includegraphics[scale=0.6]{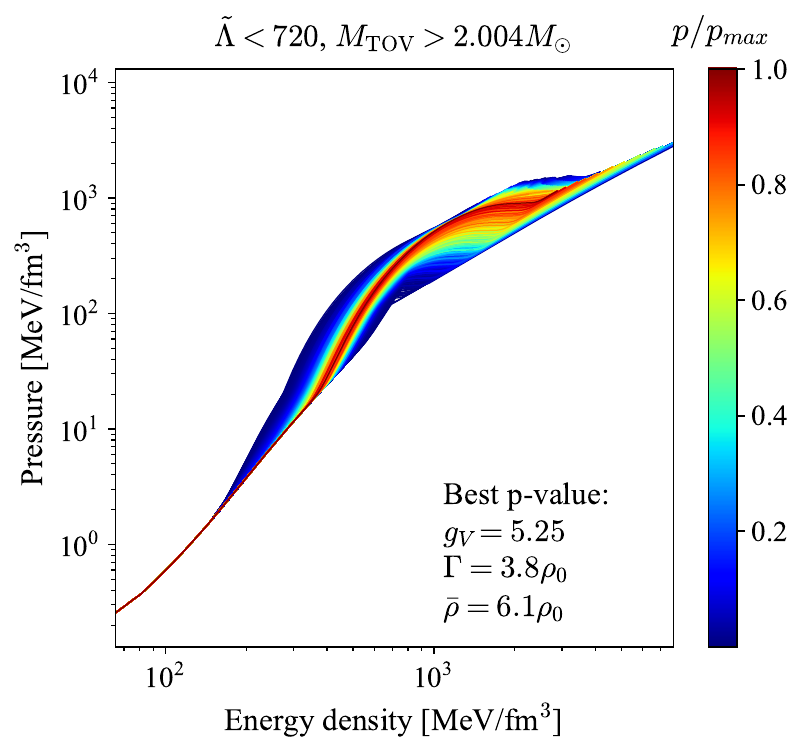}\\
        \includegraphics[scale=0.6]{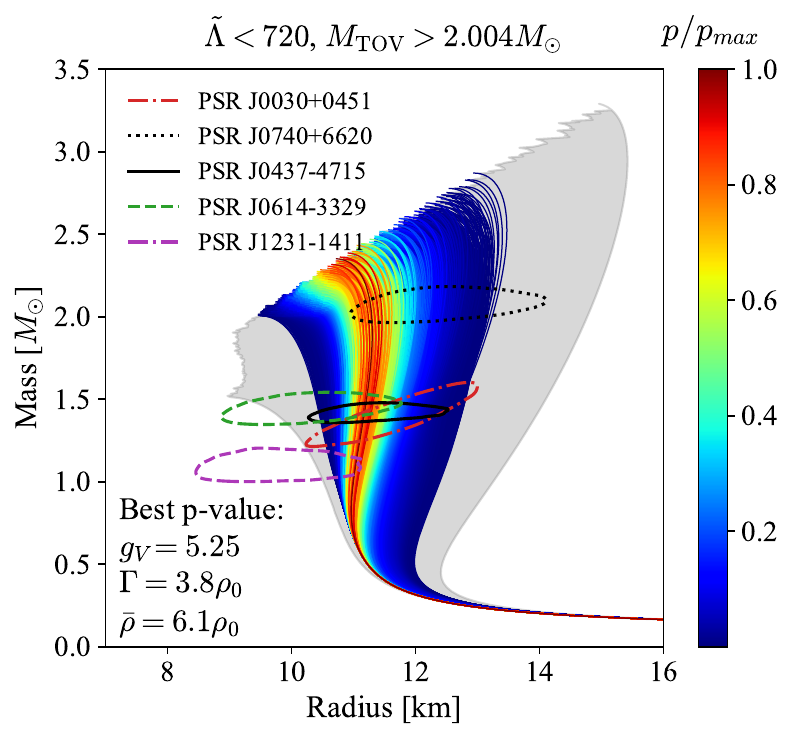}

        \caption{\textit{Upper panel:} The EoS curves, colored according to their posterior likelihood, are shown after applying the constraints
        $\tilde{\Lambda} < 720$ and $M_{\rm TOV} > 2.004\,M_{\odot}$.
        \textit{Lower panel:} Mass--radius relations corresponding to the EoS curves shown in the upper panel. The curves are colored according to
        the posterior likelihood of their corresponding EoSs. The grey shaded background indicates the region spanned by the prior, defined by the envelope of the corresponding family of curves. The most probable parameter set is indicated in each panel.}
        \label{fig:eos_mr_Mlow}
    \end{minipage}
    \hfill
    \begin{minipage}[t]{0.49\textwidth}
        \centering

        \includegraphics[scale=0.6]{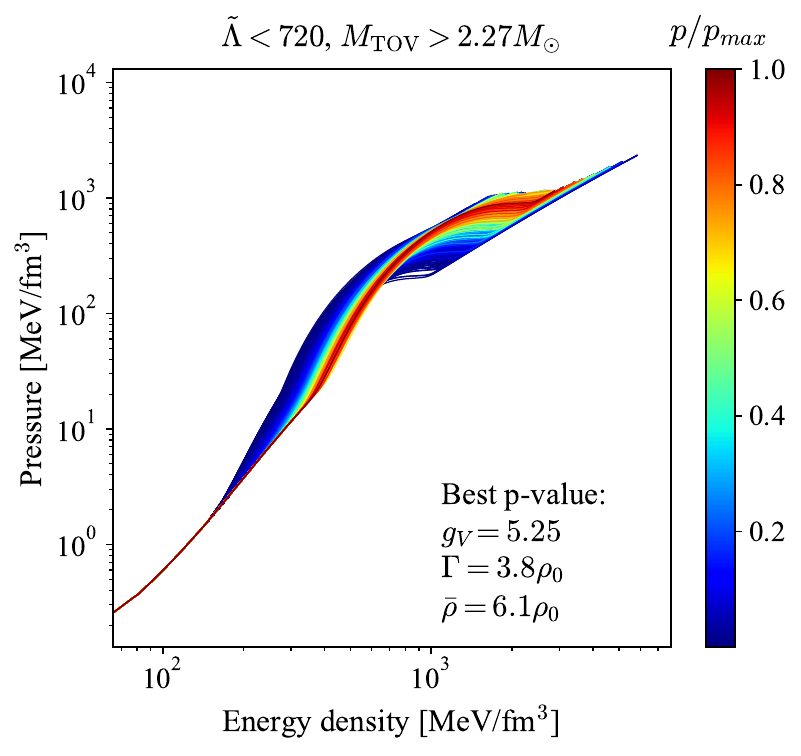}\\
        \includegraphics[scale=0.6]{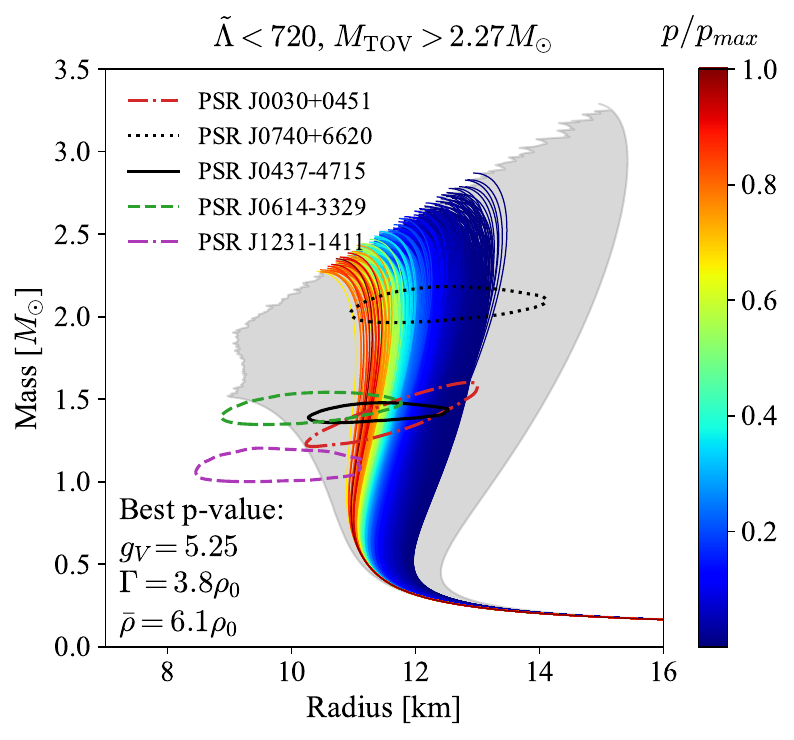}

        \caption{\textit{Upper panel:} The EoS curves, colored according to their posterior likelihood, are shown after applying the constraints $\tilde{\Lambda} < 720$ and $M_{\rm TOV} > 2.27\,M_{\odot}$. \textit{Lower panel:} Mass--radius relations corresponding to the EoS curves shown in the upper panel. The curves are colored according to the posterior likelihood of their corresponding EoSs. The grey shaded background indicates the region spanned by the prior, defined by the envelope of the corresponding family of curves. The most probable parameter set is indicated in each panel.}
        \label{fig:eos_mr_Mbw}
    \end{minipage}

\end{figure*}

Figures~\ref{fig:eos_mr_Mlow} and~\ref{fig:eos_mr_Mbw} summarize the Bayesian results obtained for the eLSM+GMSR(BSK16) hybrid EoS. Each EoS and its corresponding mass--radius sequence are
colored according to their posterior probability, with warmer colors indicating parameterizations that are more strongly supported by the combined GW170817~\cite{LIGOScientific:2018hze} and NICER observations~\cite{Vinciguerra:2023qxq,Salmi:2024aum,Mauviard:2025dmd,Choudhury:2024xbk,Qi:2025mpn}. The two figures differ only in the adopted maximum-mass constraint: Fig.~\ref{fig:eos_mr_Mlow} uses the hard constraints $\tilde{\Lambda} < 720$ and the $M_{\rm TOV} > 2.004\,M_{\odot}$, whereas Fig.~\ref{fig:eos_mr_Mbw} applies the more stringent constraint $M_{\rm TOV} > 2.27\,M_{\odot}$ together with $\tilde{\Lambda} < 720$.

For visual comparison, we also show the $1\sigma$ credible contours of the NICER measurements in these mass--radius figures. To avoid any ambiguity, we emphasize that these contours are used only for visualization, while the Bayesian analysis incorporates the full posterior distributions of the corresponding NICER measurements.

The posterior weighting considerably reduces the range of viable EoSs. Although all displayed parameterizations satisfy the theoretical requirements discussed in Sec.~\ref{sec:4}, the astrophysical observations clearly favor a relatively narrow subset of them. The highest-probability solutions correspond to EoSs that undergo a smooth hadron--quark crossover extending over a broad density
interval at relatively high average density.

The corresponding mass--radius relations demonstrate that the favored parameterizations are simultaneously consistent with all NICER-based mass--radius constraints discussed above, including the independent reanalysis of the PSR J1231--1411 observations~\cite{Vinciguerra:2023qxq,Salmi:2024aum,Mauviard:2025dmd,Choudhury:2024xbk,Qi:2025mpn}. The highest-posterior sequences pass through the $1\sigma$ confidence
regions of the measured NSs while remaining consistent with the remaining observational constraints imposed during the Bayesian analysis.

A remarkable feature of the analysis is the robustness of the inferred hybrid EoS. Keeping the tidal deformability constraint
$\tilde{\Lambda}<720$ fixed, we repeated the Bayesian analysis with either the conservative $M_{\rm TOV}>2.004\,M_\odot$ criterion or the more restrictive
$M_{\rm TOV}>2.27\,M_\odot$ constraint leads to identical highest-probability parameterizations. For both choices of the maximum-mass constraint, the most probable parameterization corresponds to
$g_V=5.25$, $\Gamma=3.8\,\rho_0$, and $\bar{\rho}=6.1\,\rho_0$, indicating that the observational data consistently favor a broad transition occurring at relatively high baryon densities.

To quantify the sensitivity of the inferred crossover parameters to the treatment of PSR J1231$-$1411, we repeated the Bayesian analysis using the alternative J1231$-$1411 posterior from Ref.~\cite{Salmi:2024bss} and also after excluding this source from the NICER likelihood. Excluding J1231$-$1411 leaves the highest-posterior parameterization unchanged at $g_V=5.25$, $\bar{\rho}=6.1\,\rho_0$, and $\Gamma=3.8\,\rho_0$. In contrast, using the alternative J1231$-$1411 analysis, which favors larger stellar radii, shifts the highest-posterior parameterization to $g_V=6.0$, $\bar{\rho}=7.4\,\rho_0$, and $\Gamma=6.4\,\rho_0$. Thus, our fiducial result is robust against the complete exclusion of J1231$-$1411, while the alternative posterior leads to a noticeable shift of the preferred crossover parameters.

\begin{figure*}[t]
    \centering

    \begin{minipage}[t]{0.49\textwidth}
        \centering

        \includegraphics[scale=0.37]{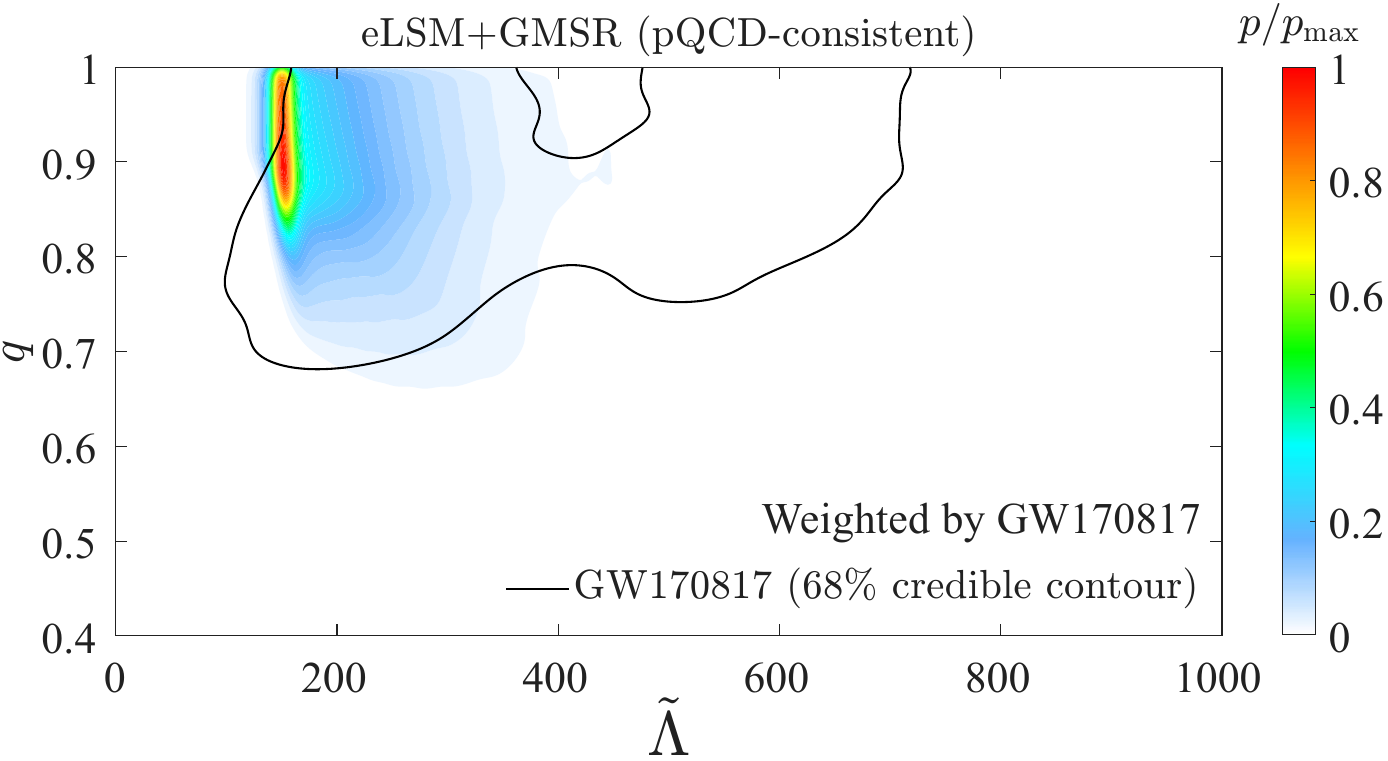}\\
        \includegraphics[scale=0.37]{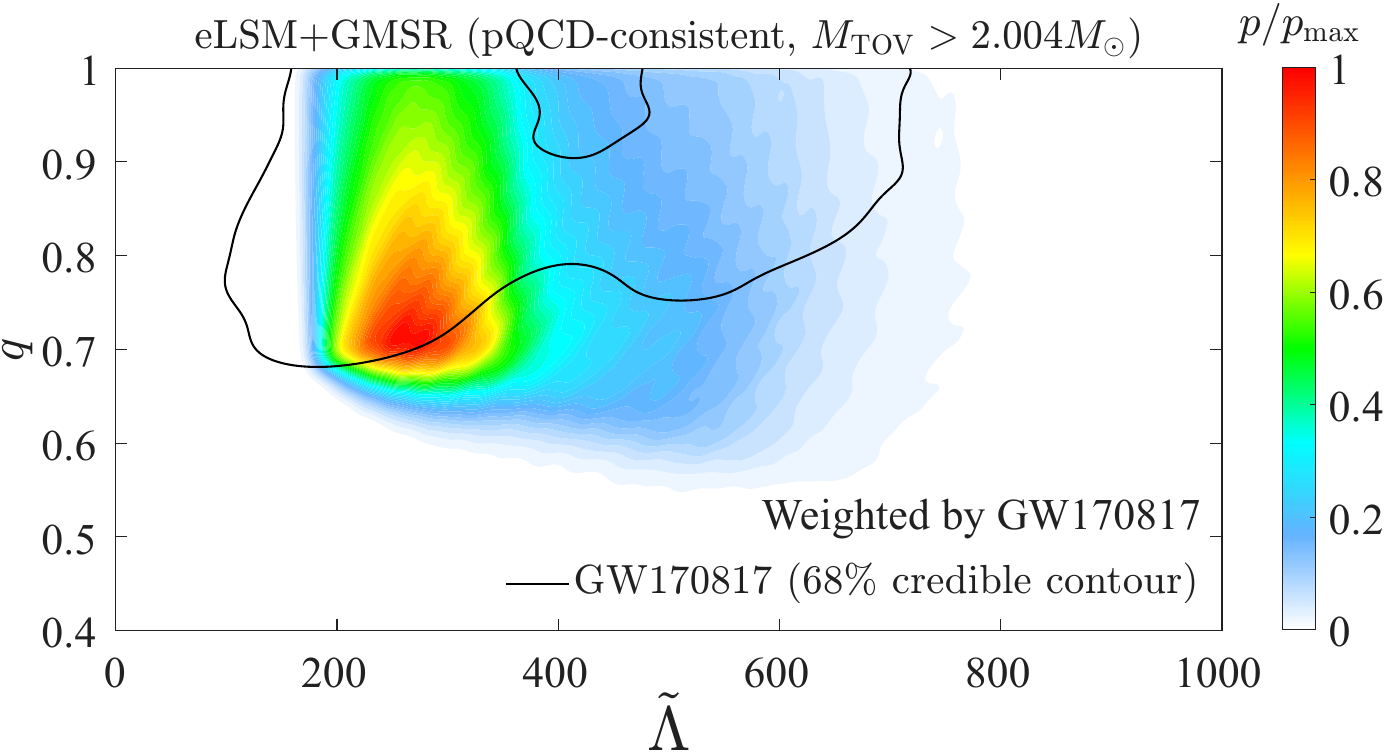}\\
        \includegraphics[scale=0.37]{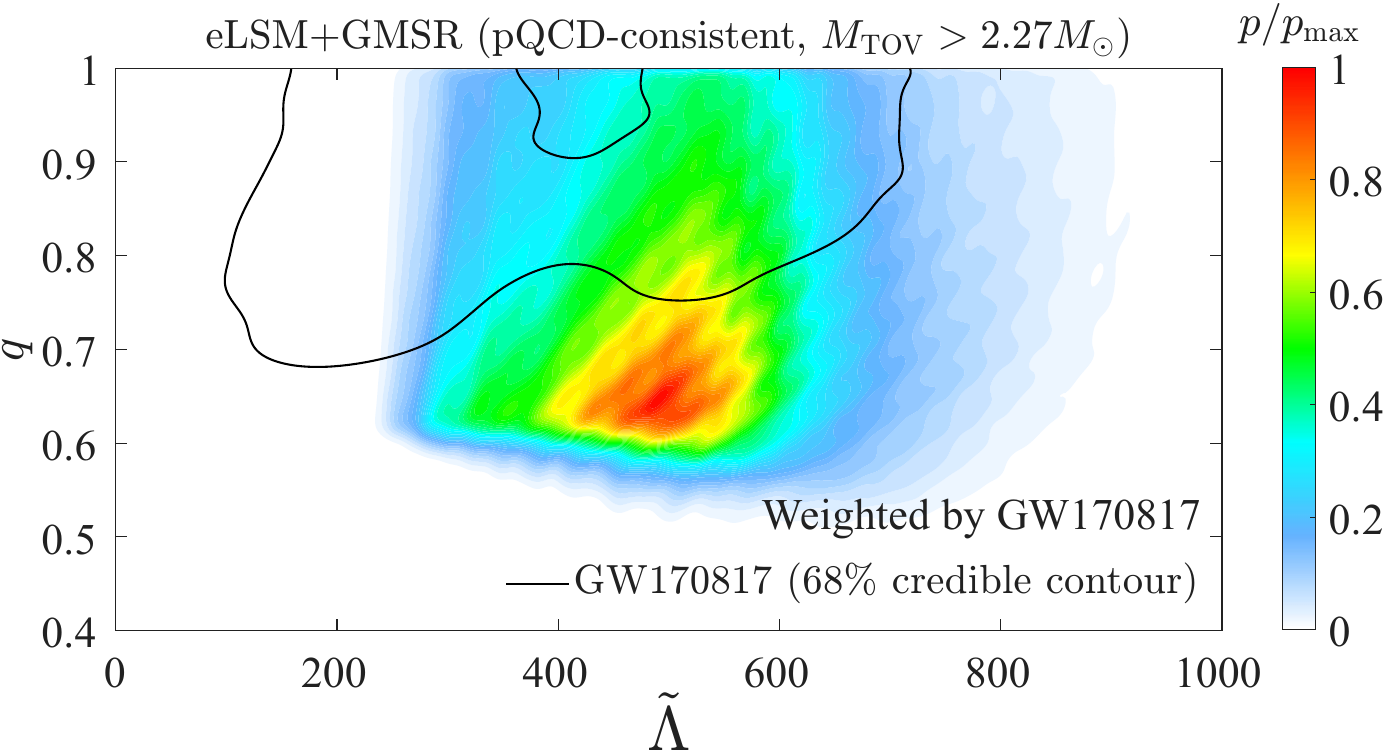}

        \caption{Posterior-weighted normalized theoretical density in the $\tilde{\Lambda}$--$q$ plane obtained using the published GW170817 posterior. The upper, middle, and lower panels correspond to the pQCD-consistent prior set, the subset satisfying $M_{\rm TOV}>2.004\,M_\odot$, and the subset satisfying $M_{\rm TOV}>2.27\,M_\odot$, respectively. The colored background shows a kernel density estimate of the posterior-weighted theoretical curves, while the black contour denotes the 68\% credible region of the published GW170817 posterior. No hard upper bound on $\tilde{\Lambda}$ is imposed in order to avoid reusing the GW170817 information already incorporated through the posterior weights.}
        \label{fig:lambda_q_GW}
    \end{minipage}
    \hfill
    \begin{minipage}[t]{0.49\textwidth}
        \centering

        \includegraphics[scale=0.37]{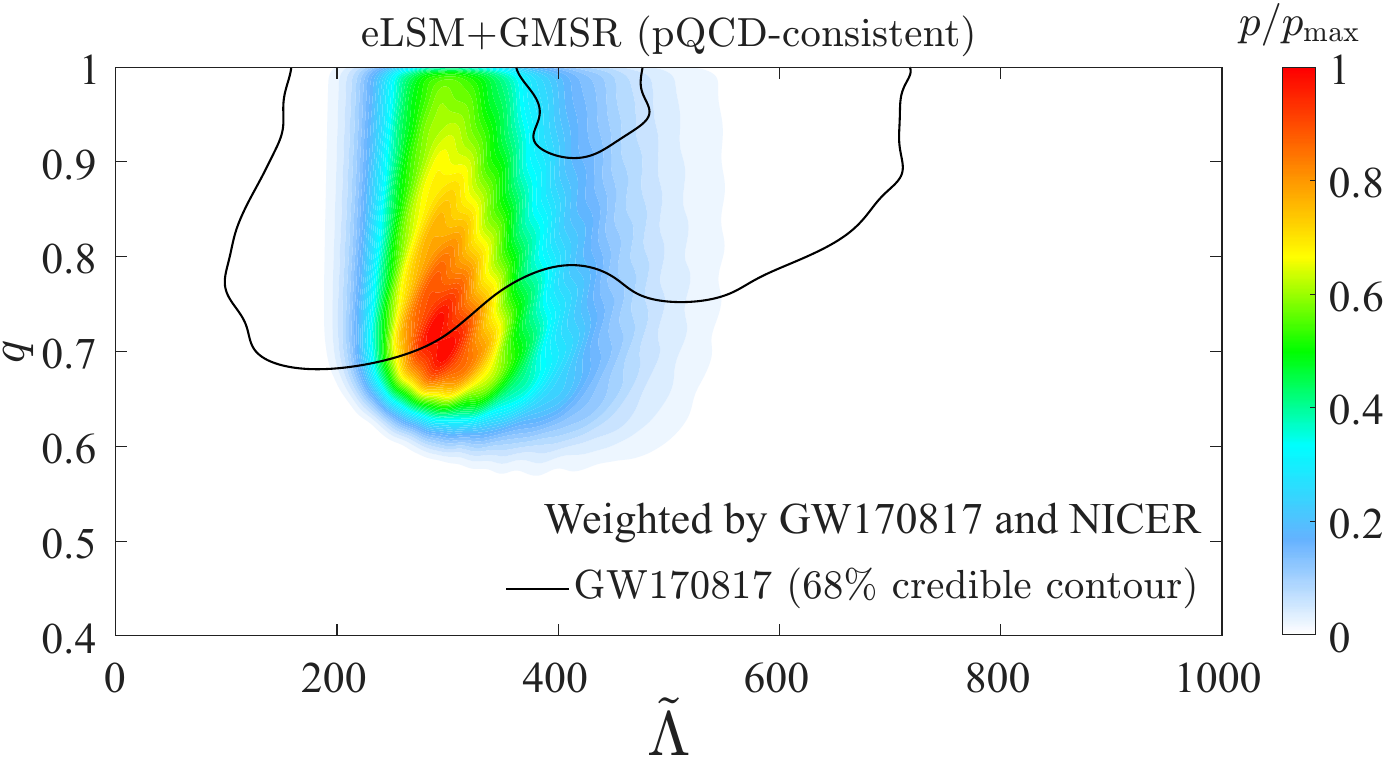}\\
        \includegraphics[scale=0.37]{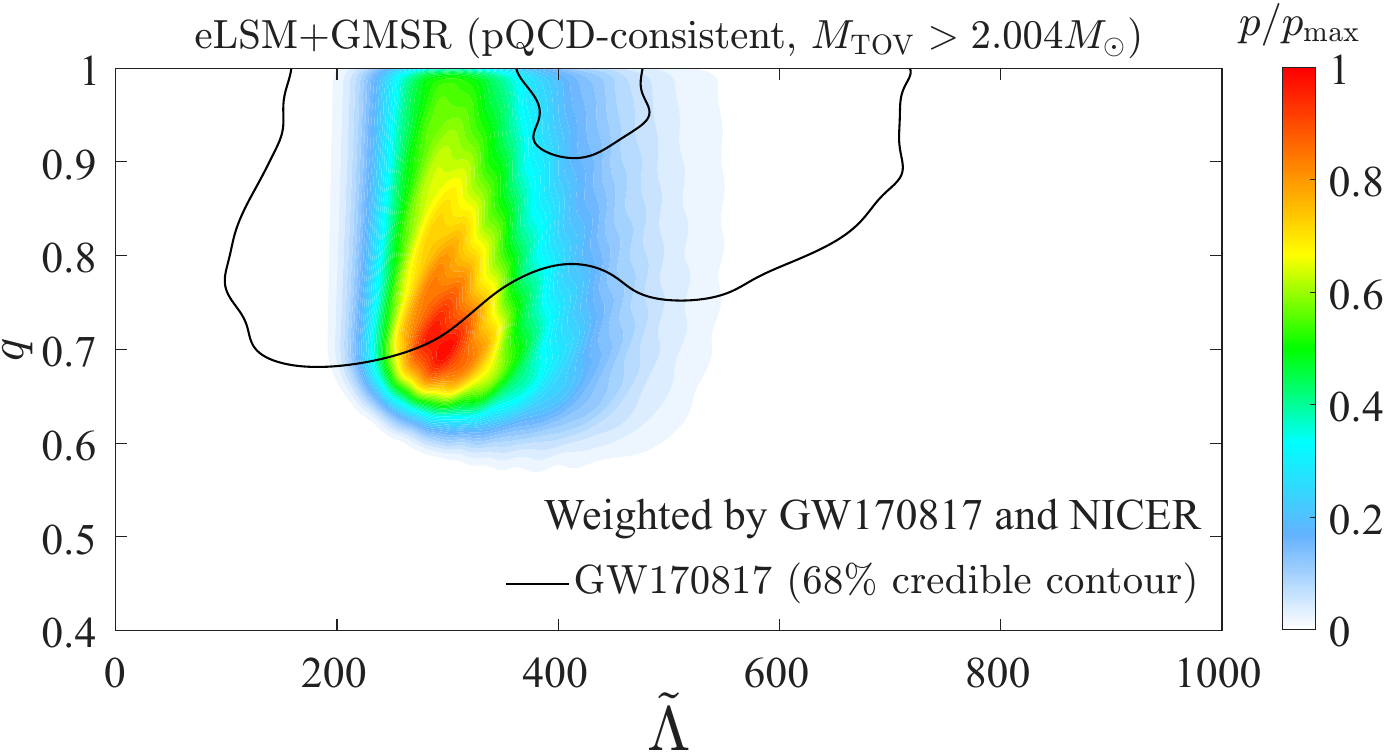}\\
        \includegraphics[scale=0.37]{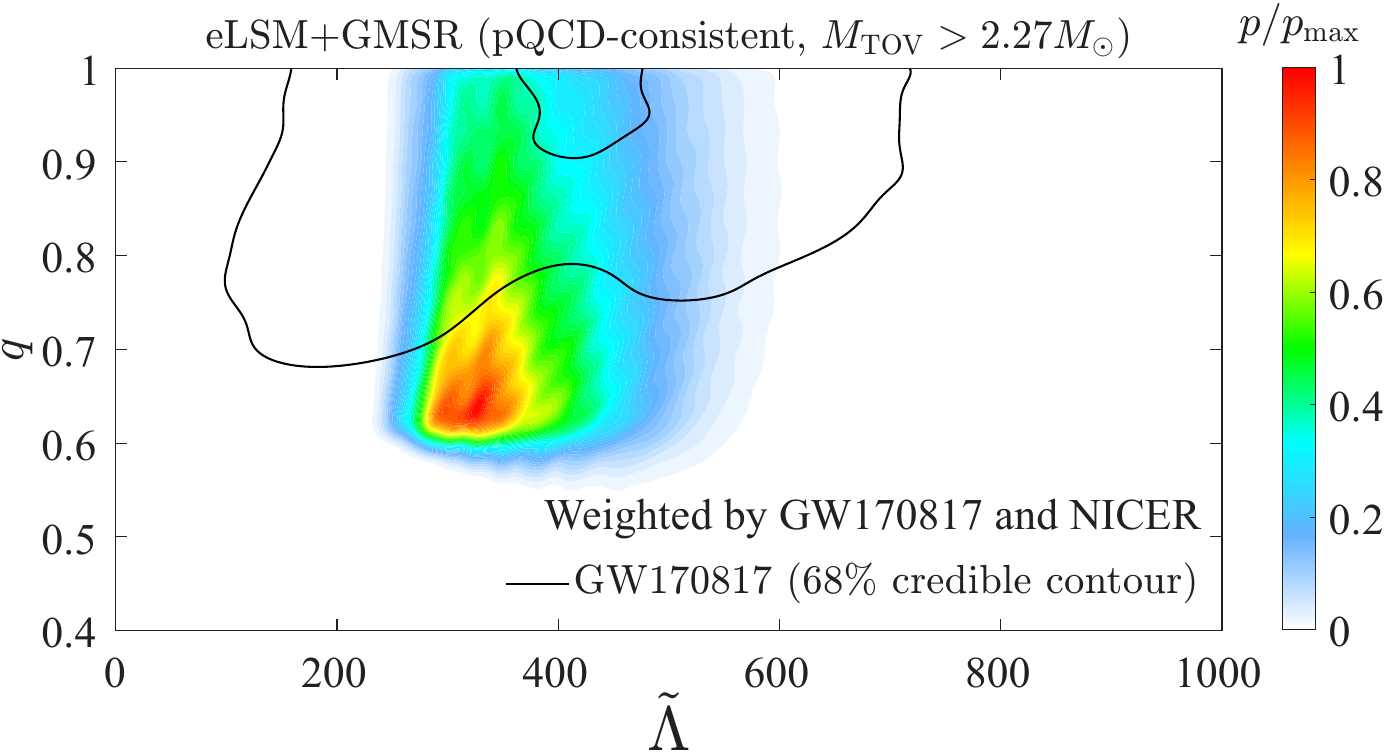}

        \caption{Posterior-weighted normalized theoretical density in the $\tilde{\Lambda}$--$q$ plane obtained using the combined published GW170817 and NICER posterior distributions. The three panels
        correspond to the same subsets as in Fig.~\ref{fig:lambda_q_GW}. Compared with the GW170817-only weighting, the inclusion of the NICER posterior produces a more concentrated theoretical density and substantially reduces the dependence on the adopted lower bound for $M_{\rm TOV}$.}
        \label{fig:lambda_q_GWNICER}
    \end{minipage}

\end{figure*}

Figure~\ref{fig:lambda_q_GW} shows the posterior-weighted normalized theoretical density in the $\tilde{\Lambda}$--$q$ plane obtained using the published GW170817 posterior~\cite{LIGOScientific:2018hze} for weighting. The three panels correspond to the pQCD-consistent prior set without an additional maximum-mass cut and after imposing the constraints $M_{\rm TOV}>2.004\,M_\odot$ and $M_{\rm TOV}>2.27\,M_\odot$, respectively. For each surviving parameterization, the corresponding $\tilde{\Lambda}(q)$ relation was calculated and weighted according to the published GW170817 posterior~\cite{LIGOScientific:2018hze} evaluated at the theoretical prediction. The colored background represents a kernel density estimate (KDE) constructed from the ensemble of these posterior-weighted theoretical curves, with warmer colors indicating regions of higher posterior density. To facilitate a direct comparison between the $\tilde{\Lambda}$--$q$ panels, the same KDE bandwidth is used throughout. It was determined automatically from the full prior sample weighted by the GW170817 posterior and subsequently kept fixed for all other selections and weighting schemes. In each panel, the density is normalized to its maximum value, such that the comparison emphasizes changes in the shape of the posterior-weighted theoretical density. No hard upper bound on $\tilde{\Lambda}$ is imposed, since the GW170817 information is already incorporated through the posterior weights; applying an additional cut $\tilde{\Lambda}<720$ would partially reuse the same observational information and artificially restrict the theoretical density toward lower tidal deformabilities. In all three cases, the regions of highest theoretical density remain consistent with the published 68\% credible region of the GW170817 analysis, although the additional constraint $M_{\rm TOV}>2.27\,M_\odot$ visibly shifts the preferred region of the theoretical density within the $\tilde{\Lambda}$--$q$ plane.

Figure~\ref{fig:lambda_q_GWNICER} presents the analogous posterior-weighted normalized theoretical density in the $\tilde{\Lambda}$--$q$ plane obtained using the combined published GW170817~\cite{LIGOScientific:2018hze} and NICER~\cite{Vinciguerra:2023qxq,Salmi:2024aum,Mauviard:2025dmd,Choudhury:2024xbk,Qi:2025mpn} posterior distributions. In this case, the weight assigned to each theoretical curve is proportional to the product of the corresponding GW170817 and NICER posterior values. In the GW170817-only analysis, the adopted lower bound on $M_{\rm TOV}$ produces a pronounced redistribution of the posterior density. By contrast, after including the NICER posterior, the three distributions become considerably more similar and are concentrated within a narrower range of tidal deformabilities. This indicates that the NICER posterior provides a dominant additional constraint on the hybrid-EoS parameter space, substantially reducing the sensitivity of the posterior-weighted theoretical density to the choice between the two maximum-mass thresholds.

We also tested the sensitivity of the NICER-weighted distributions in the $\tilde{\Lambda}$--$q$ plane to the treatment of J1231$-$1411. Excluding this pulsar produces only minor changes, while using the alternative J1231$-$1411 posterior from Ref.~\cite{Salmi:2024bss} leads to somewhat broader distributions and shifts the preferred region toward larger $\tilde{\Lambda}$, without extending beyond the $\tilde{\Lambda}$ range covered by the $1\sigma$ GW170817 contour. At the same time, the distribution in $q$ remains essentially unchanged. Thus, the qualitative behavior of the distributions is not driven by the fiducial J1231$-$1411 analysis. The corresponding distributions obtained by excluding PSR~J1231$-$1411 and by adopting the alternative posterior of Ref.~\cite{Salmi:2024bss} are presented in Appendix~\ref{app:J1231}.

\subsection{Maximum NS mass}

To investigate the implications of the hybrid EoS for the maximum NS mass, we calculated the posterior probability
distribution of $M_{\rm TOV}$ using the published GW170817~\cite{LIGOScientific:2018hze} and NICER~\cite{Vinciguerra:2023qxq,Salmi:2024aum,Mauviard:2025dmd,Choudhury:2024xbk,Qi:2025mpn} posterior distributions as weighting functions. Unlike the analyses presented in the previous subsection, no lower bound on $M_{\rm TOV}$ was imposed in order to avoid introducing a circular bias into
the inferred posterior distribution.

\begin{figure}
    \centering
    \includegraphics[scale=0.47]{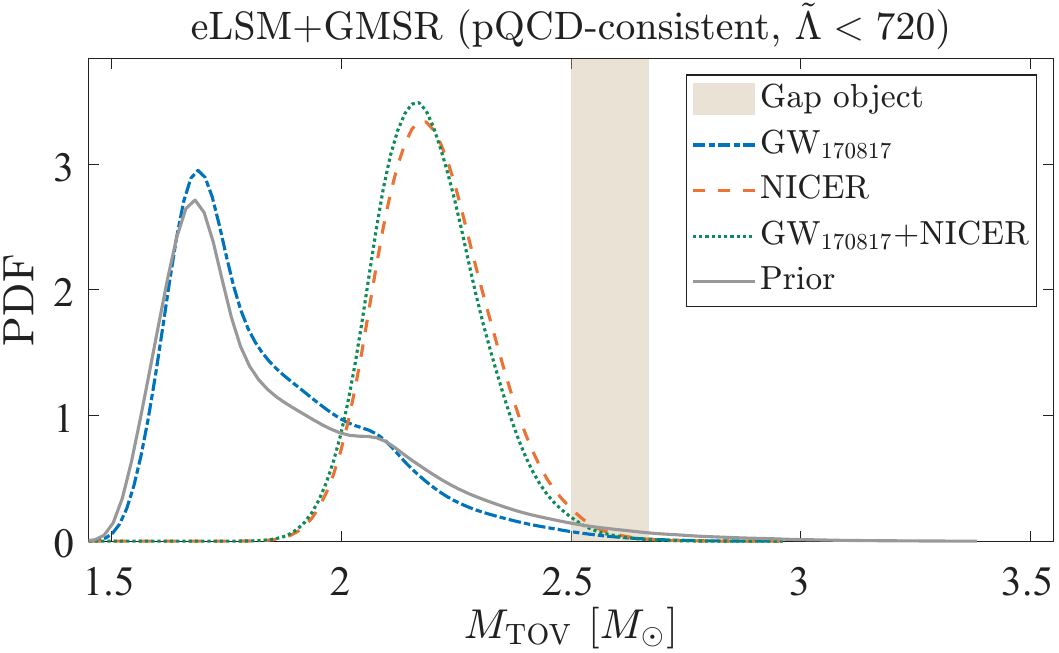}
    \includegraphics[scale=0.47]{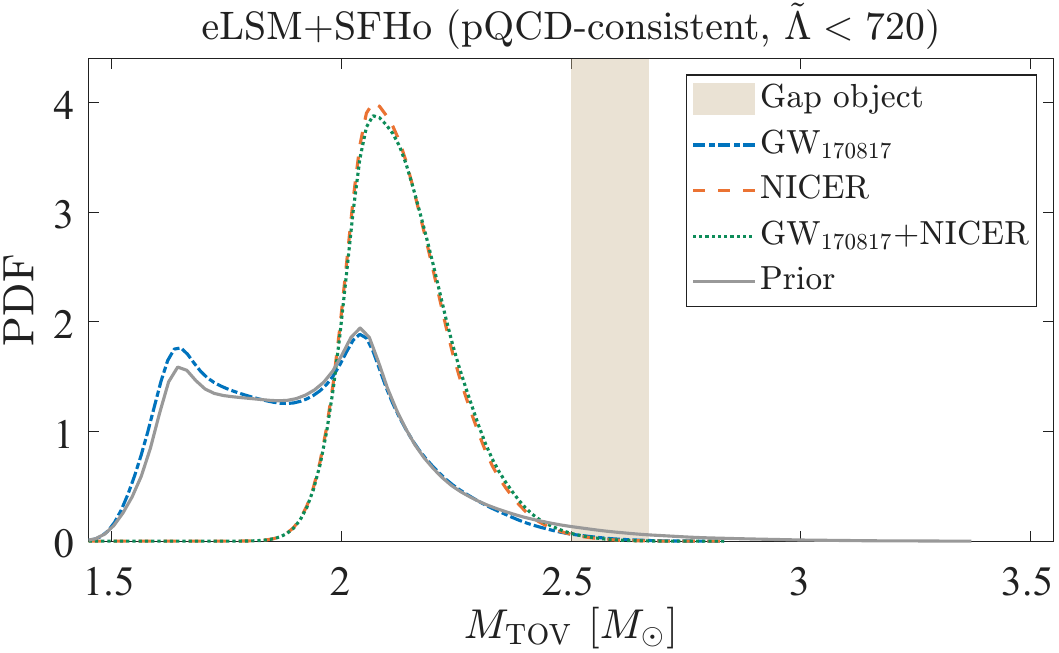}
    \caption{Posterior probability density of the maximum NS mass, $M_{\rm TOV}$, obtained from the pQCD-consistent hybrid equation-of-state parameterizations satisfying $\tilde{\Lambda}<720$. The upper panel shows results for the eLSM+GMSR(BSK16) model, while the lower panel presents the corresponding distributions for the eLSM+SFHo model. The gray solid curve denotes the prior distribution, whereas the blue, orange, and green curves show the distributions obtained by weighting the theoretical $M_{\rm TOV}$ samples according to the published GW170817 posterior, the published NICER posterior distributions, and their combination, respectively. The shaded vertical band indicates the mass range of the GW190814 secondary ("Gap object"). In both models, the combined posterior assigns only a negligible probability to maximum masses within the gap-object region, indicating that a NS interpretation of the secondary component of GW190814 is strongly disfavored.}
    \label{fig:mtov}
\end{figure}

Figure~\ref{fig:mtov} compares the prior distribution with the posterior distributions obtained using the GW170817 measurement alone, the NICER data alone, and their combination. The upper panel corresponds to the eLSM+GMSR(BSK16) hybrid EoS, while the lower panel shows the corresponding results obtained with the eLSM+SFHo model. While the prior and the GW170817 measurement permits a broad range of maximum masses, the NICER data substantially reduce the allowed interval and shift the posterior toward values higher than $2\,M_\odot$.

The theoretical posterior distribution of $M_{\rm TOV}$ is primarily determined by the NICER observations. Adding the GW170817 likelihood leaves the NICER-only posterior almost unchanged, showing that the combined constraint is dominated by the mass--radius measurements. The GW170817 data alone,
however, favor lower values of $M_{\rm TOV}$ and yield a substantially different posterior.

The inferred posterior distribution also has direct implications for the nature of the secondary compact object in GW190814, whose measured mass falls within the so-called lower mass gap~\cite{LIGOScientific:2020zkf,Biswas:2020xna}.

As shown in the upper panel of Fig.~\ref{fig:mtov}, where the GMSR(BSK16) hadronic EoS is adopted, a negligible fraction of the posterior probability extends to maximum masses comparable to that required for a stable NS interpretation of the secondary component. The overwhelming majority of the posterior is concentrated at lower maximum masses.

Within the framework of the present hybrid EoS, these results strongly disfavor a NS interpretation of the secondary component of GW190814. Although such a scenario cannot be excluded completely, it is supported by only a very small fraction of the posterior probability.

For comparison, the lower panel of Fig.~\ref{fig:mtov} shows the corresponding posterior distribution obtained using the SFHo hadronic EoS. Although the subsequent analysis focuses exclusively on the GMSR(BSK16) model, SFHo is included here as the most viable alternative among the remaining hadronic models. In this case, the posterior probability above the mass-gap region is suppressed even more strongly, indicating that the conclusion that the secondary component of GW190814 is unlikely to be a NS is not specific to the GMSR(BSK16)~model.

\subsection{Pure quark cores}\label{subseq:quarkcore}

The possible existence of pure quark cores in massive neutron stars remains an open question. Recent model-independent analyses combining astrophysical observations with \emph{ab initio} QCD constraints have suggested that the most massive stable neutron stars are expected to contain sizable quark-matter cores, provided that the speed of sound does not strongly exceed the conformal limit~\cite{Annala:2019puf}. Studies based on hybrid equations of state have further shown that the formation of extended pure quark cores requires a relatively weak hadron--quark phase transition occurring at sufficiently low densities~\cite{Ferreira:2020evu}. On the other hand, crossover-based hybrid equations of state predict that the transition from hadronic to quark matter may proceed smoothly over an extended density interval, rather than through a strong first-order phase transition~\cite{Baym:2019iky}.

To assess whether pure quark cores are supported by current astrophysical observations, we calculated the posterior-weighted distribution of the quark-core mass and radius for the maximum-mass configuration of each viable GMSR(BSK16)-based parameterization. Throughout this analysis, each parameterization is weighted according to the combined published GW170817~\cite{LIGOScientific:2018hze} and NICER~\cite{Vinciguerra:2023qxq,Salmi:2024aum,Mauviard:2025dmd,Choudhury:2024xbk,Qi:2025mpn} posterior distributions.

\begin{figure}[h!]
    \centering
    \includegraphics[scale=0.28]{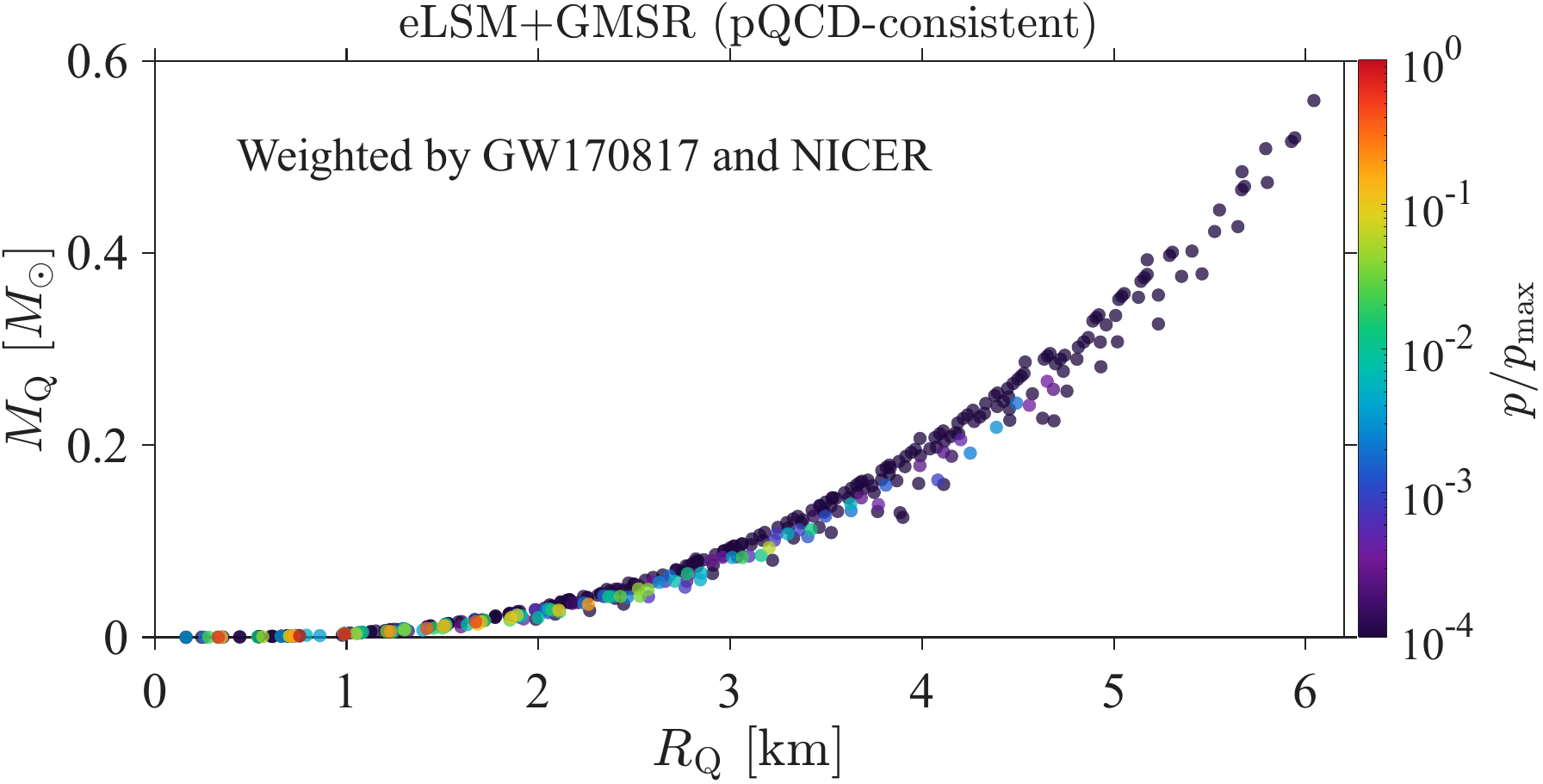}
    \includegraphics[scale=0.28]{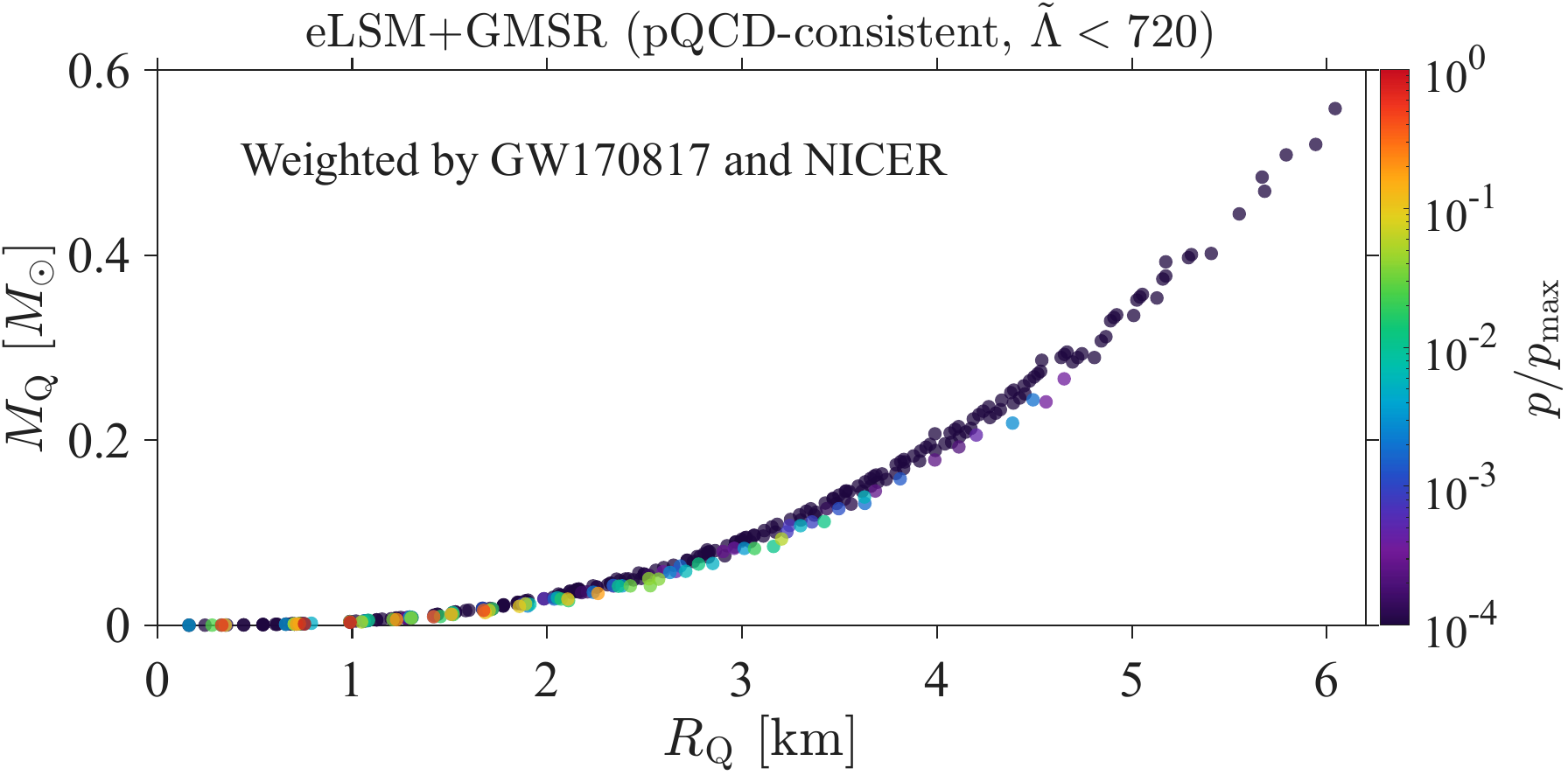}    \includegraphics[scale=0.28]{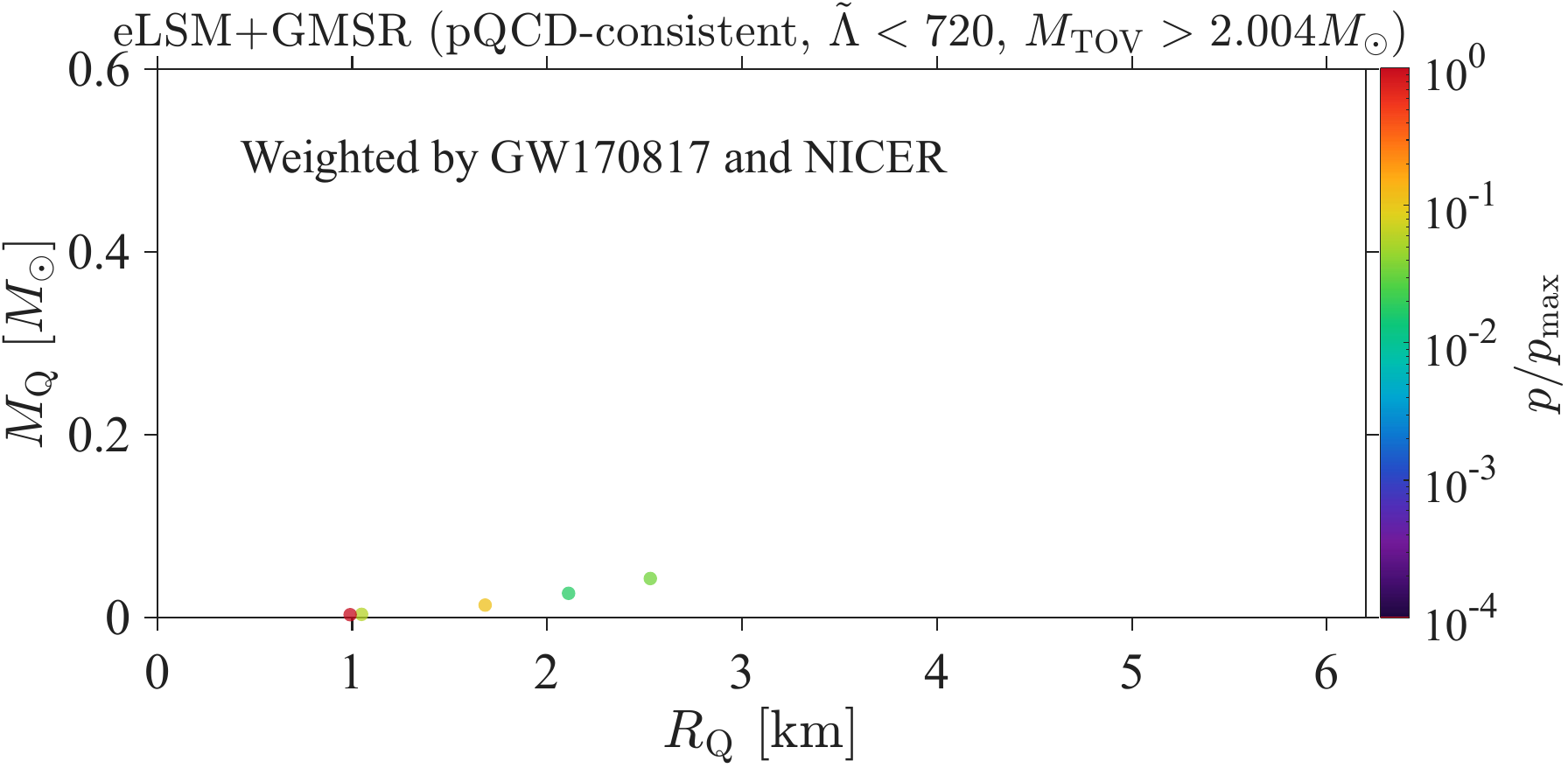}
    \includegraphics[scale=0.28]{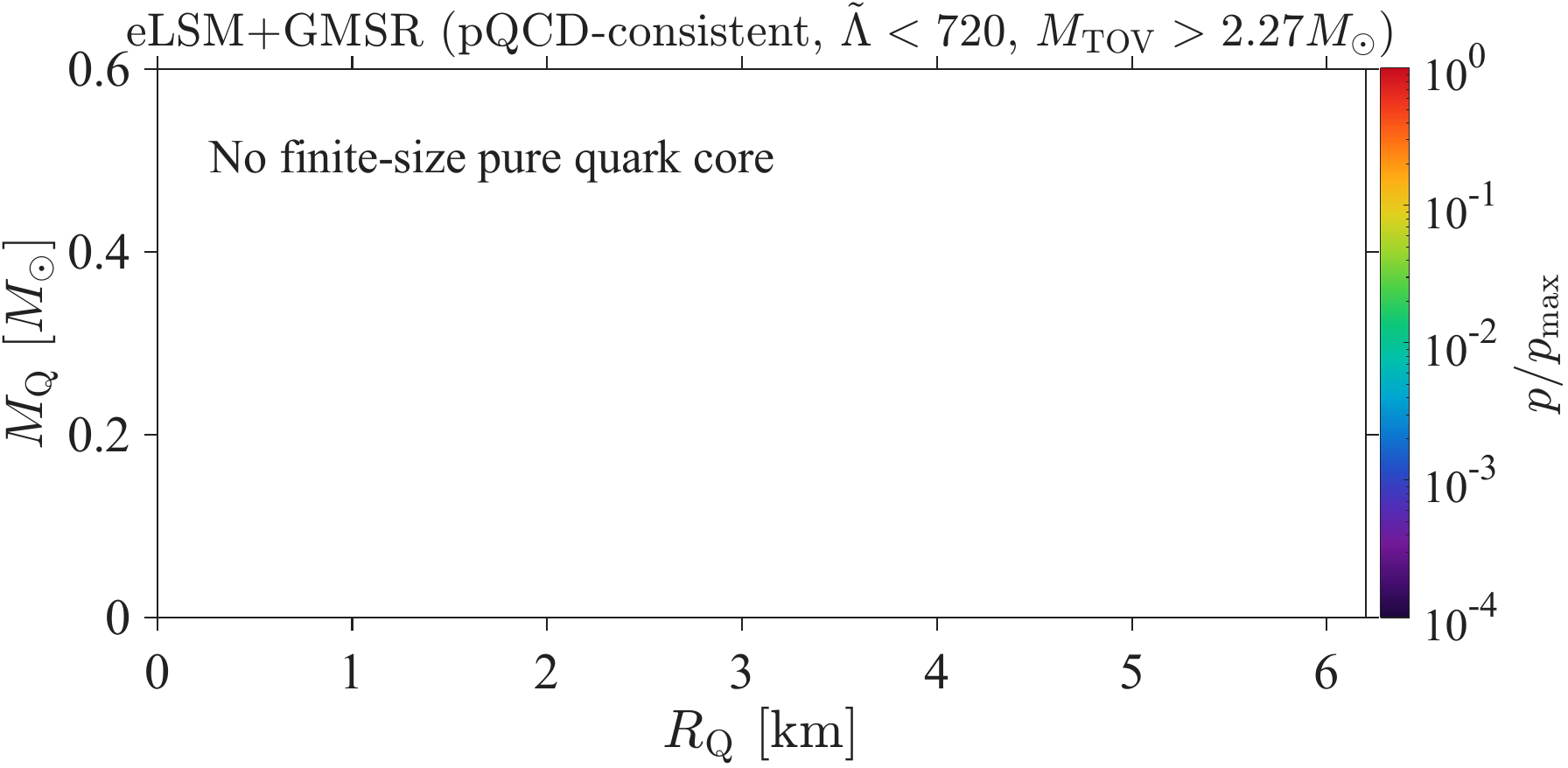}
    \caption{
    Pure quark-core mass $M_Q$ and radius $R_Q$ in the maximum-mass stellar configuration for GMSR(BSK16)-based parameterizations predicting a finite-size pure quark core. Each point represents a single hybrid-EoS parameterization and is colored according to its posterior probability, $p/p_{\rm max}$, using a logarithmic scale. The top panel shows the complete prior set, satisfying the stability, causality, and pQCD consistency conditions. The second panel additionally imposes the tidal-deformability constraint $\tilde{\Lambda}<720$, the third panel also requires $M_{\rm TOV}>2.004\,M_\odot$, while in the bottom panel this maximum-mass constraint is replaced by the stricter requirement $M_{\rm TOV}>2.27\,M_\odot$.
    }
    \label{fig:quarkcore}
\end{figure}

Figure~\ref{fig:quarkcore} shows the quark-core mass $M_Q$ and radius $R_Q$ of the maximum-mass configuration for those parameterizations that predict a finite-size pure quark core. Each point corresponds to a single hybrid-EoS parameterization and is colored according to its posterior probability normalized to the maximum posterior probability of the corresponding sample. The upper panel corresponds to the complete prior set, which satisfies the stability, causality, and pQCD consistency conditions. The middle panel additionally imposes the tidal-deformability constraint $\tilde{\Lambda}<720$, while the lower panel also includes the maximum-mass requirement $M_{\rm TOV}>2.004\,M_\odot$. Parameterizations for which the maximum-mass configuration contains no finite-size pure quark core are not displayed.

The quantitative effect of these successive constraints is summarized in Table~\ref{tab:quarkcore}. Among the 11400 parameterizations with nonzero posterior weight in the prior set, 413 (3.62\%) predict a finite-size pure quark core. However, their combined posterior weight amounts to only 0.0096\% of the total posterior probability. Thus, even before imposing the additional observational hard cuts, parameterizations containing finite-size pure quark cores are strongly disfavored by the combined GW170817 and NICER likelihoods.

The tidal-deformability constraint has only a minor effect on this result. After imposing $\tilde{\Lambda}<720$, 341 of the 10866 remaining parameterizations (3.14\%) contain a finite-size pure quark core, while their combined posterior weight remains nearly unchanged at 0.0093\%. This is also apparent in
Fig.~\ref{fig:quarkcore}, where the distribution of the surviving finite-core configurations in the $M_Q$--$R_Q$ plane changes only slightly.

The situation changes qualitatively once the maximum-mass requirement $M_{\rm TOV}>2.004\,M_\odot$ is imposed. Although 4667 parameterizations with nonzero posterior remain, only five of them predict a finite-size pure quark core in its maximum-mass configuration. Consistent with this finding, none of the 1258 parameterizations surviving the more stringent $M_{\rm TOV}>2.27\,M_\odot$ constraint contains a finite-size pure quark core. Consequently, the lower panel of Fig.~\ref{fig:quarkcore} contains no points. The maximum-mass constraint is therefore the condition that ultimately eliminates finite-size pure quark cores from the viable parameter set.

These results show that the current astrophysical likelihoods already assign an extremely small posterior weight to parameterizations predicting finite-sized pure quark cores. Imposing the additional $M_{\rm TOV}>2.004,M_\odot$ constraint nearly eliminates such configurations, while the more stringent $M_{\rm TOV}>2.27,M_\odot$ constraint removes them entirely within the present hybrid-EoS framework. In contrast, the $\tilde{\Lambda}<720$ constraint has only a minor effect on either their relative abundance or their posterior weight.

To assess the dependence of this conclusion on the choice of the hadronic EoS, we repeated the pure-quark-core analysis for the SFHo-based models on the coarse parameter grid introduced in Sec.~\ref{sec:4c}. Of the 2775 parameterizations satisfying the theoretical prior, 1900 carry nonzero posterior weight, among which 115 (6.05\%) contain a finite-size pure quark core, accounting for only 0.0081\% of the total posterior weight. After imposing $\tilde{\Lambda}<720$, 85 out of the 1620 parameterizations with nonzero posterior weight (5.25\%) retain a pure quark core, with a combined posterior weight of 0.0070\%. Most importantly, none of the 711 parameterizations that additionally satisfy $M_{\rm TOV}>2.004\,M_\odot$ contains a finite-size pure quark core. Thus, although the detailed results presented in Table~\ref{tab:quarkcore} are based on the refined GMSR(BSK16) grid, the SFHo coarse-grid analysis leads to the same qualitative conclusion regarding the absence of pure quark cores once the tidal-deformability and maximum-mass constraints are imposed. This common behavior reflects a more general tendency of the crossover construction: simultaneously satisfying the tidal-deformability and maximum-mass constraints favors sufficiently broad crossover regions, which in turn suppress the formation of finite-size pure quark cores in stable NS configurations. We do not perform the same comparison for DD2 and DNS(CMF), since these constructions already show substantial tension with the combined tidal-deformability and maximum-mass observational hard cuts in the coarse-grid analysis of Sec.~\ref{sec:4c} and were therefore not retained for the subsequent detailed analysis.

The absence of pure quark cores does not imply the absence of quark matter inside NSs. On the contrary, for the GMSR(BSK16)-based models, the most probable parameterizations consistently predict an extended hadron--quark crossover region, in which hadronic and quark degrees of freedom coexist over a broad density interval. For the highest-posterior parameterization
($g_V=5.25$, $\Gamma=3.8\,\rho_0$, and $\bar{\rho}=6.1\,\rho_0$), the maximum-mass configuration has
$M_{\rm TOV}=2.351\,M_\odot$ and
$R_{\rm TOV}=10.724\,\mathrm{km}$, while no pure quark core is formed ($M_Q=0$, $R_Q=0$). Instead, a hadron--quark crossover region is already present for stellar configurations with
$M_{\rm H-Q}=2.133\,M_\odot$ and
$R_{\rm H-Q}=9.699\,\mathrm{km}$.

\begin{table*}[t]
\caption{
Number and posterior weight of GMSR(BSK16)-based parameterizations predicting a finite-size pure quark core in the maximum-mass NS configuration after the successive application of the observational constraints.
Here $N_{\rm tot}$ denotes the number of parameterizations with
nonzero posterior weight after each selection step, $N_Q$ is the
number of parameterizations within this set that predict a finite-size pure quark core, and $p_Q$ is their total normalized posterior weight.
}
\label{tab:quarkcore}
\centering
\begin{ruledtabular}
\begin{tabular}{lcccc}
Selection & $N_{\rm tot}$ & $N_Q$ & $N_Q/N_{\rm tot}$ & $p_Q$ \\
\hline
Prior
& 11400 & 413 & 3.62\% & $9.62\times10^{-5}$ \\
Prior + $\tilde{\Lambda}<720$
& 10866 & 341 & 3.14\% & $9.34\times10^{-5}$ \\
Prior + $\tilde{\Lambda}<720$ + $M_{\rm TOV}>2.004\,M_\odot$
& 4667 & 5 & 0.11\% & $2.08\times10^{-5}$ \\
Prior + $\tilde{\Lambda}<720$ + $M_{\rm TOV}>2.27\,M_\odot$
& 1258 & 0 & 0\% & 0 \\
\end{tabular}
\end{ruledtabular}
\end{table*}

\section{Discussion and comparison with recent hybrid-EoS studies}
\label{sec:6}

A closely related Bayesian crossover analysis was recently presented by Grundler and Li~\cite{Grundler:2026zus}, who coupled a hadronic meta-model to a trace-anomaly-based quark-matter parameterization through a smooth switching function. Their posterior favors a crossover centered at $\varepsilon \simeq (4\text{--}6)\,\varepsilon_0$ (where $\varepsilon_0$ denotes the energy density of symmetric nuclear matter at saturation), with a width of approximately $(0.5\text{--}1.0)\,\varepsilon_0$, broadly consistent with our preference for an extended transition at relatively high density. Their quark-matter parameterization is constructed to approach the asymptotic conformal limits $c_s^2\to1/3$ and $P/\varepsilon\to1/3$. However, no explicit matching to the finite-density pQCD pressure band is imposed; instead, consistency with pQCD is motivated by the asymptotic behavior of the parameterization.  The qualitative agreement between their preferred crossover regime and that found in the present work is therefore noteworthy despite the different treatments of the high-density QCD regime.

A complementary analysis was presented by Albino et al.~\cite{Albino:2025puc}, who investigated hybrid-star ensembles based on two different quark-matter descriptions, the NJL and MFTQCD (Mean-Field-Theory QCD) models, using a Maxwell construction for the hadron--quark phase transition. The comparatively large canonical radii obtained for the NJL-based ensembles of that work should not be interpreted primarily as a direct signature of NJL quark matter. In most accepted NJL parameterizations, the Maxwell transition occurs only above the central density of a $1.4\,M_\odot$ star, with a median transition mass of $M_{\rm trans}\simeq1.7\,M_\odot$. The canonical radius is therefore determined predominantly by the hadronic RMF sector. Requiring an appreciable quark core together with a sufficiently large maximum mass
preferentially selects stiff hadronic parameterizations, characterized by enhanced vector repulsion and large positive higher-order nuclear-matter coefficients. This leads to
$R_{1.4}\simeq13.3\text{--}13.7$ km and produces tension with the GW170817 tidal-deformability constraint. Thus, the large radii mainly reflect correlations induced between the assumed quark model, transition construction, and hadronic prior.

The better agreement of the MFTQCD ensemble of that work with small-radius and tidal-deformability constraints can be traced primarily to its early Maxwell transition. With a median onset density of approximately $1.4\,\rho_0$, the quark phase already controls the intermediate-density pressure relevant for canonical neutron-star radii, whereas the NJL transition generally occurs only above the central density of a $1.4\,M_\odot$ star. Subsequent vector stiffening of the MFTQCD phase permits the same EoSs to support two-solar-mass stars. This produces the observationally favorable combination of a relatively soft intermediate-density EoS and a stiff high-density EoS. Nevertheless, the inferred early onset is conditioned by the imposed transition-density range and by the requirement that stable stars contain large quark cores; it should therefore not be regarded as model-independent evidence for deconfinement near saturation density.

MFTQCD assumes current-mass, chirally symmetric quarks already at the transition, whereas NJL retains large dynamical quark masses and chiral-restoration dynamics. Consequently, the very early transition removes the EoS from the tightly constrained hadronic regime shortly above saturation density, allowing the flexible quark model to determine a large fraction of the stellar properties. The observational success of the MFTQCD ensemble therefore relies on the combination of an early transition, which softens the EoS in the density regime relevant for canonical neutron-star radii, and sufficient stiffening at higher densities to support massive neutron stars. This should not, however, be interpreted as evidence that the astrophysical data independently require such an early first-order transition.

Among the studies considered here, Ref.~\cite{Albino:2025puc} is also particularly relevant with respect to the treatment of the high-density QCD regime, as it explicitly incorporates the Gorda--Komoltsev--Kurkela pQCD likelihood~\cite{Gorda:2022jvk}, evaluated at $7\,\rho_0$.

Ref.~\cite{Ayriyan:2025rub} employs the hyperonic DD2Y-T functional as the hadronic component and obtains hybrid equations of state compatible with neutron-star observations. This does not contradict the disfavoring of the nucleonic DD2 construction found here. In Ref.~\cite{Ayriyan:2025rub}, the preferred Maxwell transition occurs already at approximately $(1\text{--}2)\rho_0$, so that the DD2Y-T sector is replaced by color-superconducting quark matter below the central densities of essentially all observed neutron stars. The resulting radii and maximum masses are therefore governed primarily by the CFL quark phase rather than by the high-density DD2Y-T behavior. Their nonlocal quark-matter model approaches the asymptotic conformal limits expected from pQCD, although no explicit matching to the finite-density pQCD pressure band is imposed.

As discussed above, the treatment of the high-density QCD regime differs substantially among these studies. Ref.~\cite{Albino:2025puc} explicitly
incorporates the Gorda--Komoltsev--Kurkela pQCD likelihood at $7\,\rho_0$, whereas Refs.~\cite{Ayriyan:2025rub,Grundler:2026zus}
incorporate the expected asymptotic behavior of QCD without explicitly imposing the finite-density pQCD pressure band. In the present work, pQCD consistency is imposed by requiring each accepted hybrid EoS to admit a thermodynamically stable and causal continuation from the endpoint of the neutron-star sequence to the pQCD constraints on pressure, baryon density, and chemical potential. This distinction is important because approaching the conformal limits $c_s^2\to1/3$ and $P/\varepsilon\to1/3$ is not, by itself, equivalent to satisfying the finite-density pQCD constraint, which also restricts the pressure and baryon density at finite chemical potential, including the associated renormalization-scale uncertainty. An EoS may therefore possess the correct asymptotic conformal behavior while remaining incompatible with the finite-density pQCD constraint.

\section{Conclusions}

In this work, we have investigated the properties of hybrid NSs using phenomenological hadron--quark crossover EoS constructed by smoothly connecting hadronic and quark matter. The quark phase was described within the extended Linear Sigma Model, while four representative hadronic EoSs were considered to account for the present theoretical uncertainties. The model parameters were constrained through Bayesian inference using current gravitational-wave and NICER observations.

Our analysis shows that the combination of present astrophysical observations strongly constrains the properties of the hadron--quark crossover. The comparison of the four hadronic baselines reveals a pronounced dependence on the underlying hadronic EoS. The GMSR(BSK16)-based construction provides the highest joint probability among the models considered, while SFHo remains viable but is less favored by the adopted joint NICER and LVC likelihood. In contrast, DD2 and DNS(CMF) encounter substantial tension with the combined tidal-deformability and maximum-mass hard cut constraints. Furthermore, the preferred parameterizations consistently favor a broad crossover occurring at relatively high densities, while narrow transition regions are strongly disfavored. The refined parameter scan and an additional high-resolution convergence test show that the location of the maximum-posterior region is stable under further grid refinement.

The maximum-posterior parameterization simultaneously passes through the 68\% credible regions of all NICER-based mass--radius constraints considered in this work~\cite{Vinciguerra:2023qxq,Salmi:2024aum,
Mauviard:2025dmd,Choudhury:2024xbk,Qi:2025mpn}, which include an independent reanalysis of PSR J1231$-$1411, demonstrating excellent overall agreement with the current NICER mass--radius constraints. Excluding PSR J1231$-$1411 from the likelihood leaves the maximum-posterior parameterization unchanged, whereas adopting the alternative J1231$-$1411~\cite{Salmi:2024bss} analysis, which favors larger radii, shifts the preferred crossover parameters toward a broader transition at higher density. Thus, the fiducial result is robust against the exclusion of this source, although it remains sensitive to the adopted J1231$-$1411 posterior. The posterior-weighted theoretical density in the $\tilde{\Lambda}$--$q$ plane remains consistent with the published GW170817 posterior~\cite{LIGOScientific:2018hze}, while the inclusion of the joint NICER posterior significantly narrows the preferred region and largely removes the sensitivity to the adopted lower bound on the maximum NS mass.

The marginalized posterior distributions provide a more direct characterization of the preferred transition region. In terms of the physical crossover boundaries, the onset density is considerably more tightly constrained than the completion density, with median values $\rho_{\rm low}=1.790^{+0.350}_{-0.399}\,\rho_0$ and $\rho_{\rm high}=11.469^{+4.103}_{-2.352}\,\rho_0$. This indicates that the present
observations favor an extended crossover whose onset occurs close to $2\,\rho_0$, while its high-density completion remains much less tightly constrained.

We also investigated the posterior distribution of the maximum NS mass without imposing a lower bound on $M_{\rm TOV}$, thus avoiding circular reasoning. The resulting posterior is largely determined by the NICER measurements and favors maximum masses consistent with the existence of the heaviest observed pulsars.
This behavior is found for both the GMSR(BSK16)- and SFHo-based hybrid constructions, despite the substantial differences in their prior maximum-mass distributions.
Although an NS interpretation of the secondary component of GW190814~\cite{LIGOScientific:2020zkf} cannot be completely excluded, the corresponding posterior probability is found to be very small.

Finally, we examined whether current observations support the existence of pure quark cores inside NSs. Although a small fraction of the theoretical prior permits finite-size pure quark cores, these configurations carry only a negligible fraction of the combined GW170817 and NICER posterior weight. The tidal-deformability constraint has only a minor effect on this conclusion, whereas imposing $M_{\rm TOV}>2.004\,M_\odot$ nearly eliminates finite-size pure quark cores in the GMSR(BSK16)-based constructions. The more stringent $M_{\rm TOV}>2.27\,M_\odot$ constraint removes them completely. The SFHo-based constructions exhibit the same qualitative tendency, with finite-size pure quark cores already excluded by the $M_{\rm TOV}>2.004\,M_\odot$ requirement. These findings suggest that the suppression of pure quark cores is not a feature specific to the GMSR(BSK16)-based hybrid constructions.
At the same time, the preferred solutions consistently contain an extended hadron--quark crossover region, suggesting that quark degrees of freedom may already be present inside massive NSs without forming a distinct pure quark phase.

Comparison with recent hybrid-EoS studies reveals both common trends and important differences in the inferred properties of hybrid stars, particularly regarding the density range of the hadron--quark transition and the possible formation of pure quark cores. These differences highlight the remaining model dependence in extracting the microscopic properties of dense matter from current astrophysical observations.

The present study was restricted to nonrotating stellar configurations. Future work will extend Bayesian analysis to rapidly rotating NSs by incorporating rotational equilibrium sequences and the associated astrophysical constraints. Such an extension will make it possible to explore the influence of rotation on the inferred hadron--quark crossover parameters and on the internal composition of NSs, providing an important step toward a more comprehensive description of compact-star matter.

\appendix
\section{Sensitivity of the $\tilde{\Lambda}$--$q$ distributions
to the treatment of PSR J1231$-$1411}
\label{app:J1231}

\begin{figure*}
    \begin{minipage}[t]{0.49\textwidth}
        \centering
        \includegraphics[scale=0.37]{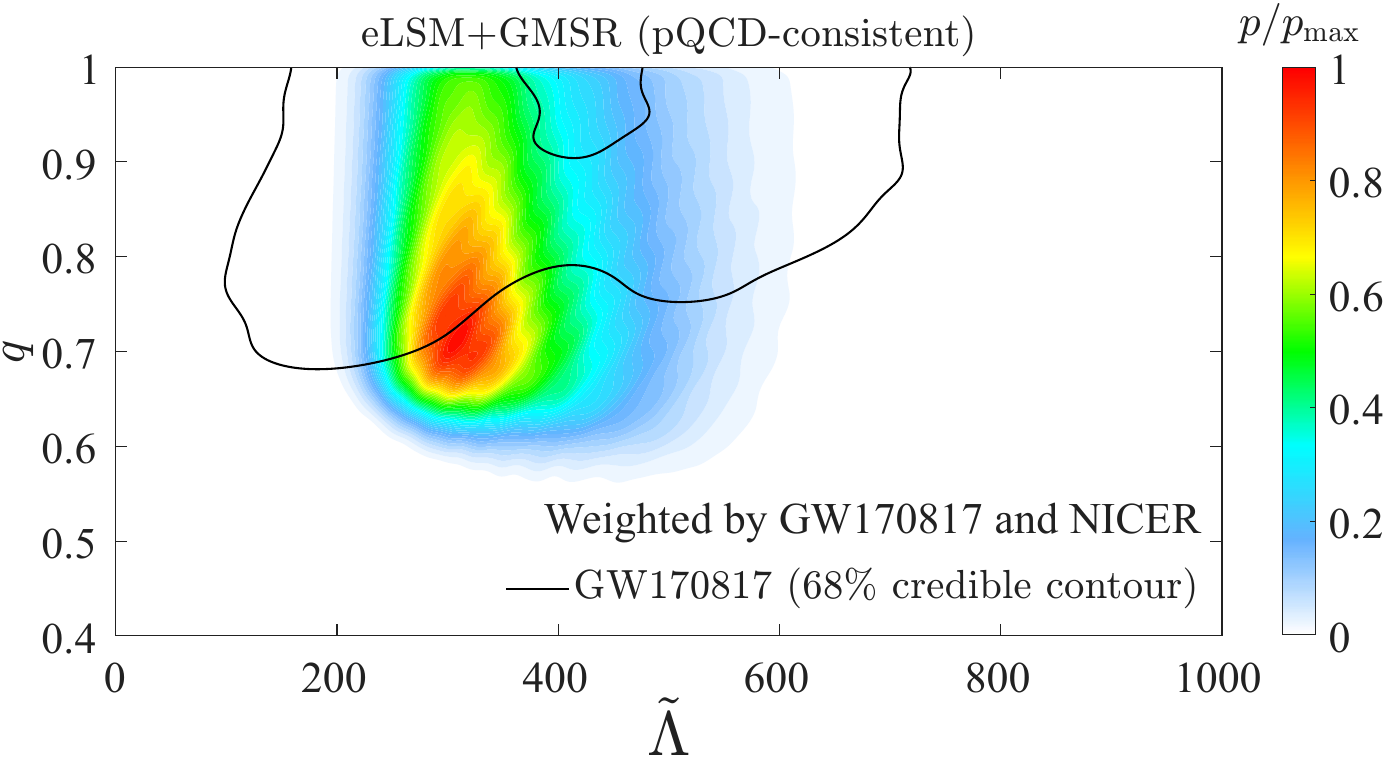}\\
        \includegraphics[scale=0.37]{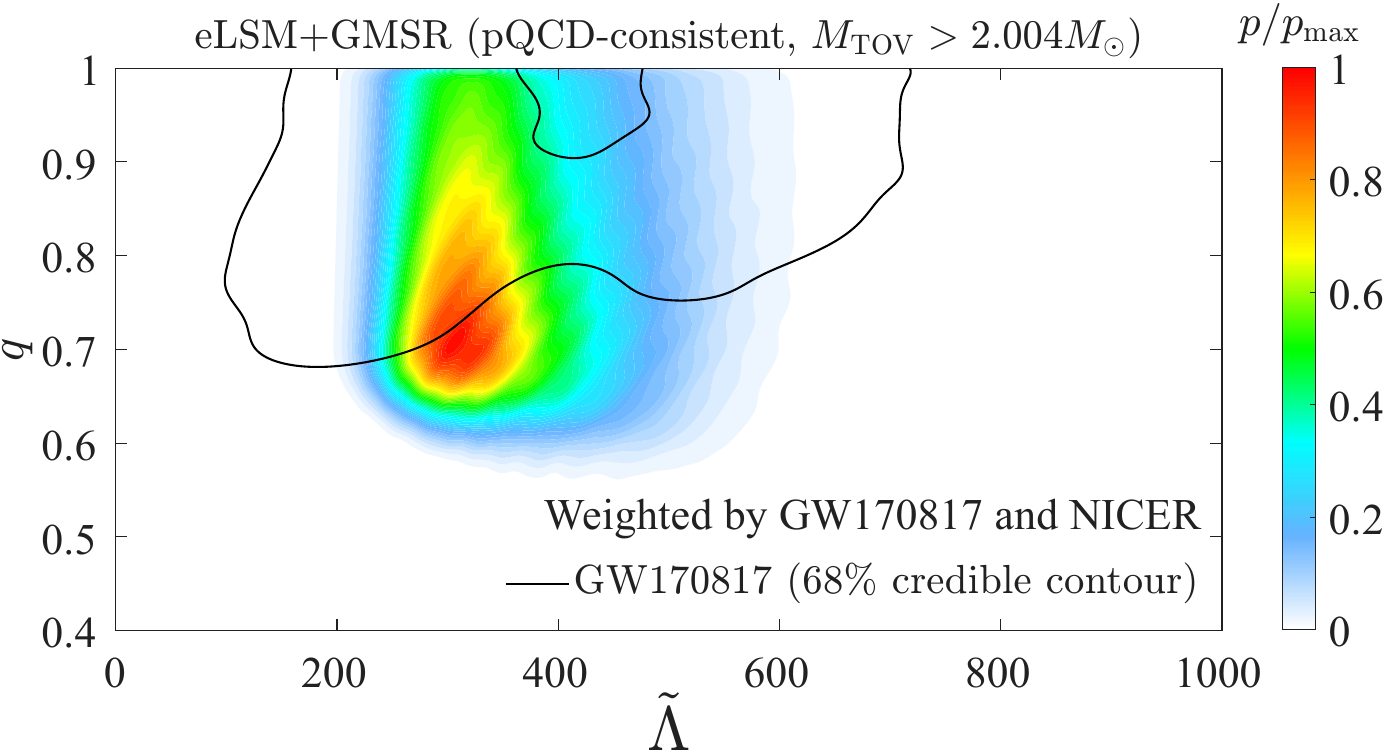}\\
        \includegraphics[scale=0.37]{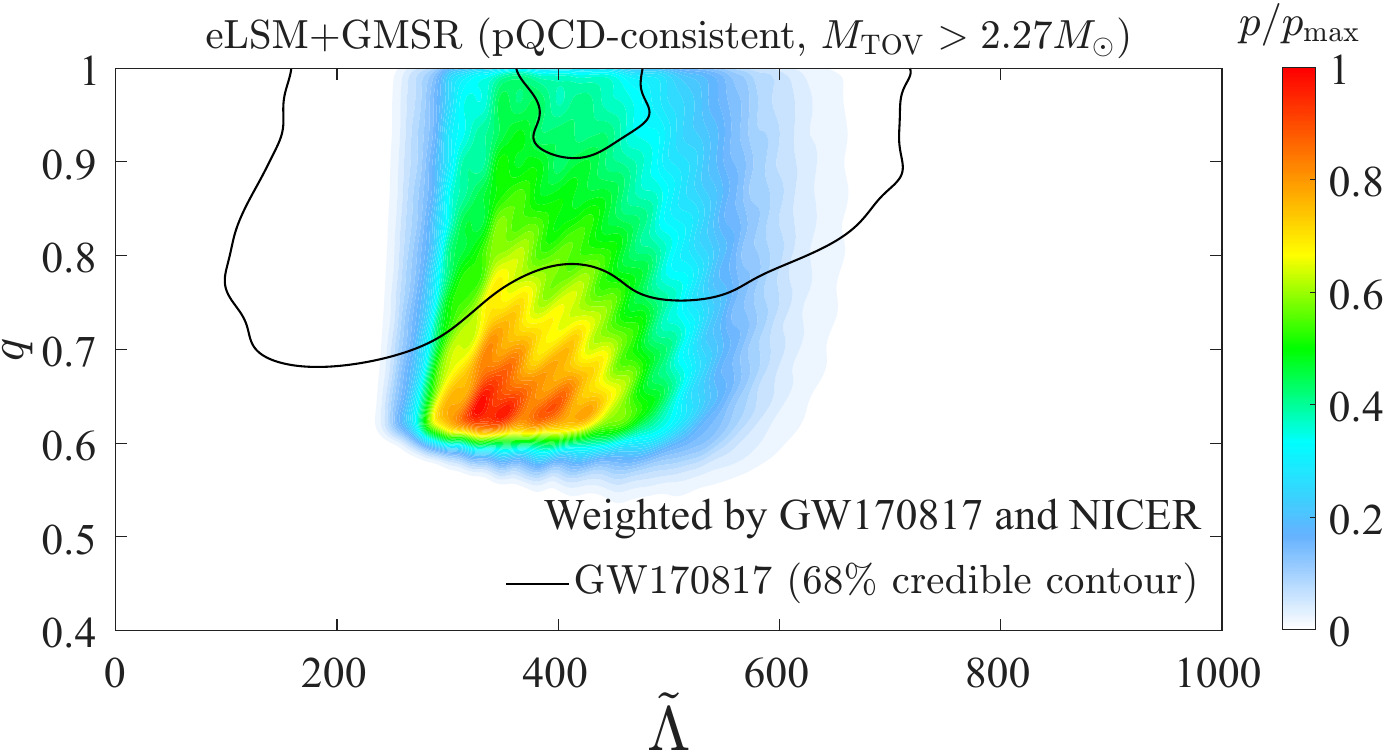}
        \caption{Posterior-weighted normalized theoretical density in the $\tilde{\Lambda}$--$q$ plane for the eLSM+GMSR(BSK16) hybrid EoS, obtained after excluding PSR~J1231$-$1411 from the NICER likelihood. The upper, middle, and lower panels correspond to the pQCD-consistent prior set, the subset satisfying $M_{\rm TOV}>2.004\,M_\odot$, and the subset satisfying $M_{\rm TOV}>2.27\,M_\odot$, respectively. The distributions are weighted by the remaining NICER posteriors and GW170817 and normalized to their respective maxima. The black curve denotes the 68\% credible contour of GW170817.}
        \label{fig:J1231_excluded}
    \end{minipage}
    \hfill
    \begin{minipage}[t]{0.49\textwidth}
        \centering
        \includegraphics[scale=0.37]{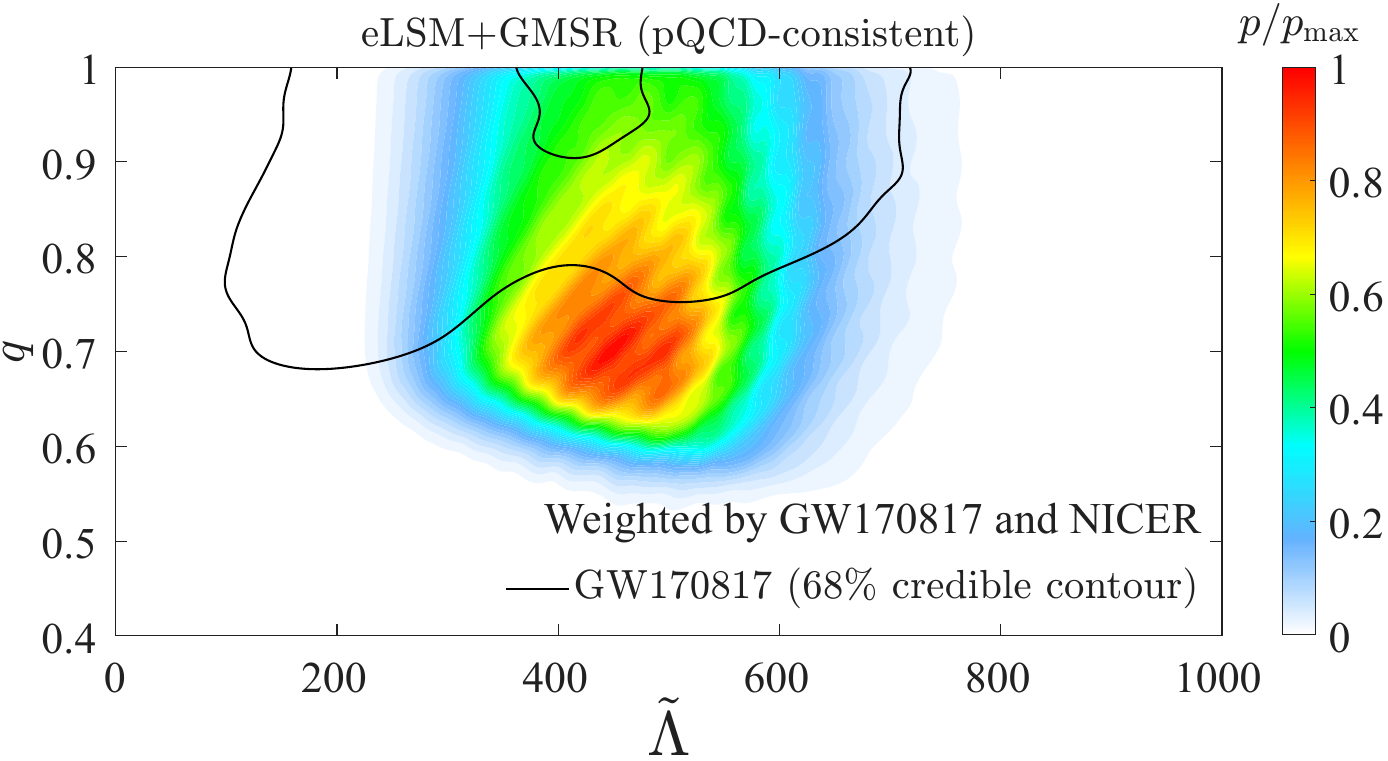}\\
        \includegraphics[scale=0.37]{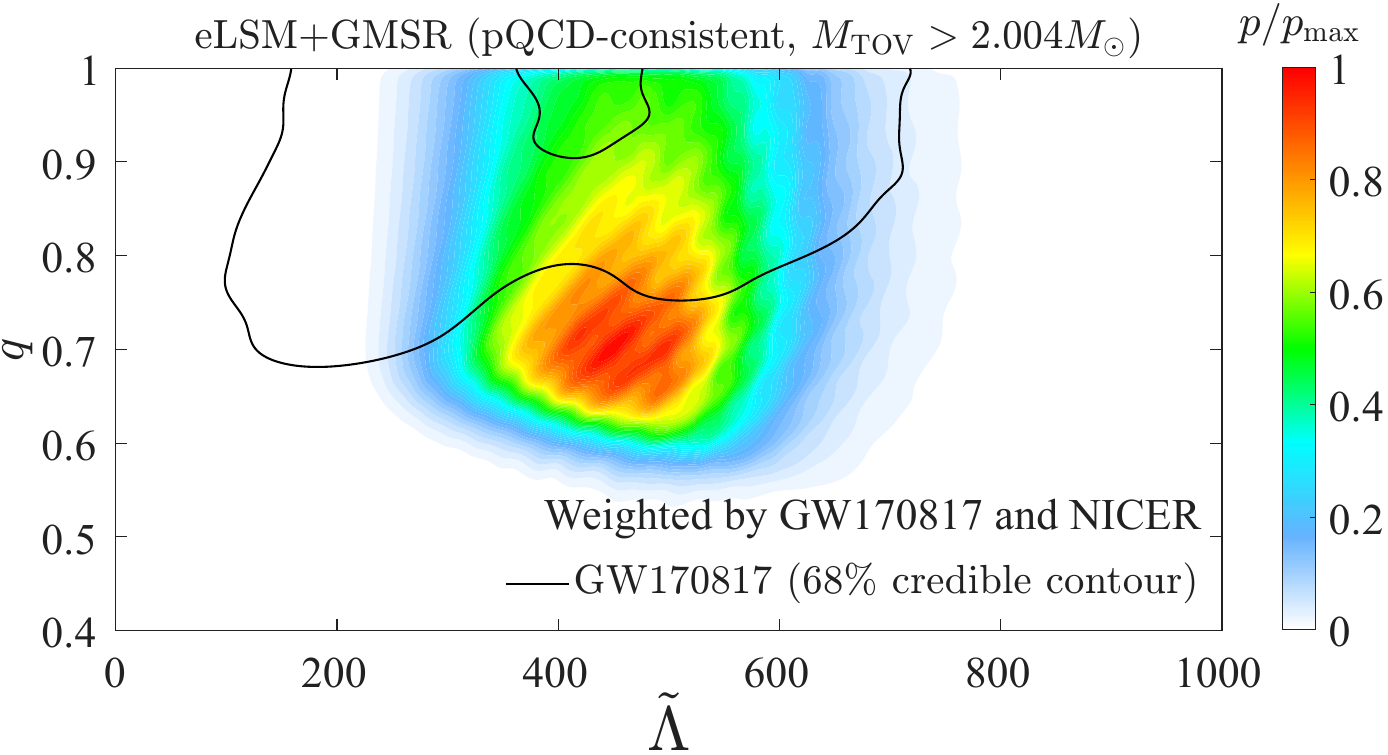}\\
        \includegraphics[scale=0.37]{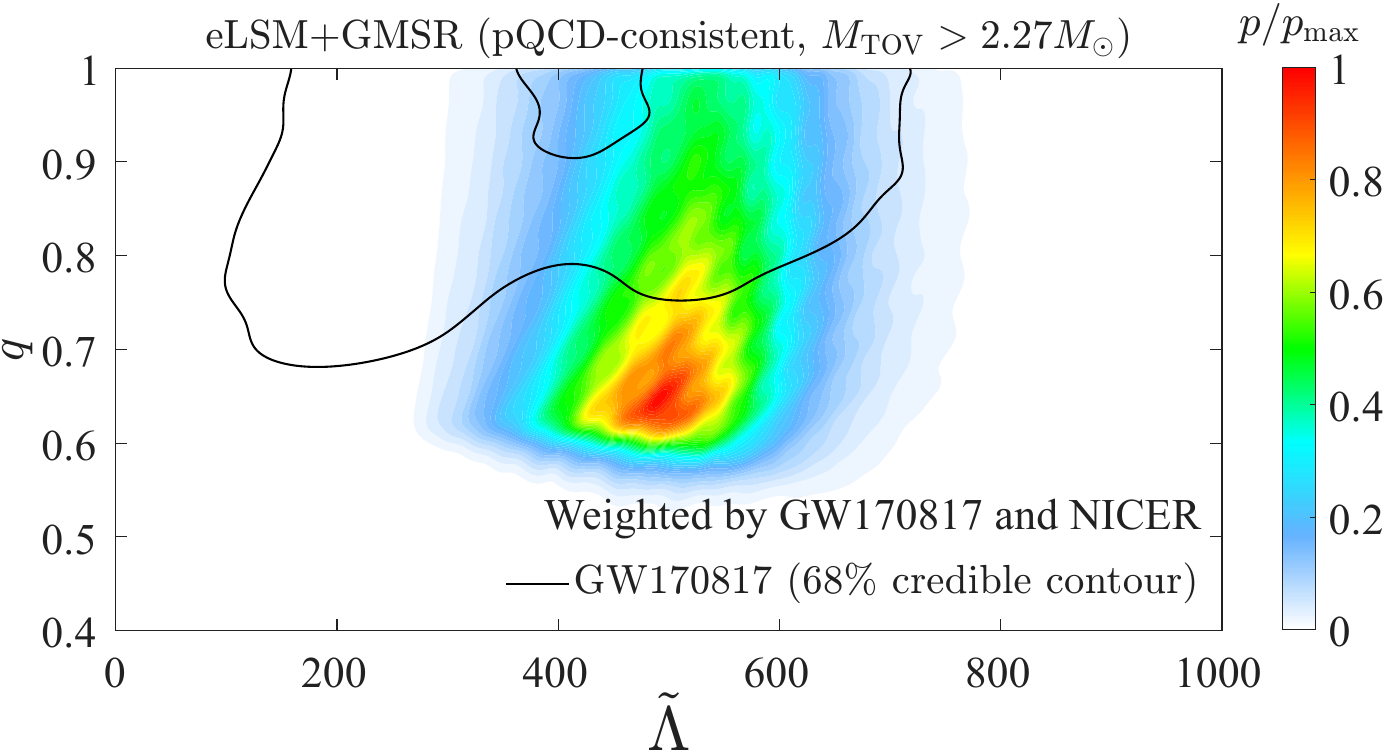}
        \caption{Same as Fig.~\ref{fig:J1231_excluded},
        but using the alternative PSR~J1231$-$1411 posterior of Salmi \textit{et al.}~\cite{Salmi:2024bss} in place of the fiducial posterior of Ref.~\cite{Qi:2025mpn}.}
        \label{fig:J1231_Salmi}
    \end{minipage}
\end{figure*}

To assess the sensitivity of the posterior-weighted $\tilde{\Lambda}$--$q$ distributions to the adopted NICER analysis of PSR~J1231$-$1411, we repeated the calculation for the GMSR(BSK16)-based hybrid EoSs using two alternative treatments of this source. In the first test, PSR~J1231$-$1411 was excluded from the combined NICER likelihood, with the corresponding distributions presented in Fig.~\ref{fig:J1231_excluded}. In the second, the fiducial posterior of Ref.~\cite{Qi:2025mpn} was replaced by the alternative posterior of Ref.~\cite{Salmi:2024bss}, with the corresponding distributions presented in Fig.~\ref{fig:J1231_Salmi}.

These figures clearly illustrate that the
qualitative behavior of the $\tilde{\Lambda}$--$q$
distributions is not driven by the fiducial treatment of PSR~J1231$-$1411.

\section*{Acknowledgements}
This work was supported by the Hungarian OTKA fund K138277.
G. L. Kasza acknowledges support from the Research Excellence Programme of the Hungarian University of Agriculture and Life Sciences.


\bibliographystyle{apsrev4-2}
\bibliography{refs}

\end{document}